\documentstyle[12pt,aaspp4]{article}
\begin{document}
\def\Ha{H$\alpha$ }
\def\Pa{Pa$\alpha$ }
\def\Hans{H$\alpha$}
\def\Pans{Pa$\alpha$}
\def\hst{{\it HST }} 
\def\gtorder{\mathrel{\raise.3ex\hbox{$>$}\mkern-14mu
             \lower0.6ex\hbox{$\sim$}}}
\def\ltorder{\mathrel{\raise.3ex\hbox{$<$}\mkern-14mu
             \lower0.6ex\hbox{$\sim$}}}
\def\asec{^{\prime\prime}}
\def\asecp{^{\prime\prime}_.}
\def\degree{^{\circ}}
\def\nod{\nodata}

\title{An Ultraviolet through Infrared Look at Star Formation and  Super Star Clusters
in Two Circumnuclear Starburst Rings\footnote 
{Based on observations with the {\it Hubble Space
Telescope} which is operated by AURA, Inc., under NASA contract NAS
5-26555.}}

\author{Dan Maoz\footnote{ Permanent Address: School 
of Physics \& Astronomy and Wise Observatory,
Tel-Aviv University, Tel-Aviv 69978, Israel.}}
\affil{Department of Astronomy, Columbia University,
550 W. 120th St., New York, NY 10027}

\authoremail{dani@wise.tau.ac.il}

\author{Aaron J. Barth}
\affil{Harvard-Smithsonian Center for Astrophysics,
60 Garden Street, Cambridge, MA 02138}
\authoremail{abarth@cfa.harvard.edu}

\author{Luis C. Ho}
\affil{The Observatories of the Carnegie Institution of Washington,
 813 Santa Barbara St., Pasadena, 
CA 91101}
\authoremail{lho@ociw.edu}

\author{Amiel Sternberg}
\affil{School of Physics \& Astronomy,
Tel-Aviv University, Tel-Aviv 69978, Israel}
\authoremail{amiel@wise.tau.ac.il}

\and

\author{Alexei V. Filippenko}
\affil{Department of Astronomy,
University of California, 
Berkeley, CA 94720-3411}
\authoremail{alex@astro.berkeley.edu}

\begin{abstract}
We present broad-band ($U, V, I,$ and $H$) and narrow-band
(\Hans+[N~II] and \Pans) images of the circumnuclear starburst rings
in two nearby spiral galaxies, NGC~1512 and NGC~5248, obtained
with the WFPC2 and NICMOS cameras on {\it HST}. Combined
with previously published ultraviolet (UV) \hst images at 2300 \AA, these data 
provide a particularly wide wavelength range with which to
study the properties of the stellar populations, the gas, and
the dust in the rings.
The young star clusters and the line-emitting gas
 have different spatial distributions, with some large (50-pc scale) 
line-emitting regions that have little associated continuum emission, but
a \Pa equivalent width indicating a few-Myr-old embedded stellar population.
The observed \Hans/\Pa\ intensity ratios suggest
the gas is mixed with dust, making it effective
 at obscuring some of the young clusters. 
  We identify
the major (about 500 in each galaxy) compact continuum sources
(super star clusters and individual stars) and analyze their spectral
energy distributions (SEDs) from 0.2 $\mu$m to 1.6 $\mu$m
by fitting them with  a grid of spectral synthesis models with 
a range of ages and dust extinction.
Most of the visible clusters are only mildly
reddened, with $A_V=0$ to 1 mag,
 suggesting that the processes that clear out the gas and dust of the
stellar birth clouds are efficient and fast. The patchiness of the 
dust distribution makes it difficult to reliably
estimate  the star formation rate,
based on UV continuum slope or hydrogen emission-line ratios, 
in starbursts such as these. 
The cluster SEDs are consistent with a range in ages, from 1~Myr to
300~Myr, but with only a minority older
than a few tens of Myr. We point out an age bias, the result
of the steep luminosity function of the clusters combined with the fading
of clusters as they age, which causes
young clusters to be over-represented at any luminosity. Accounting for this
bias, the fraction of old clusters is consistent with
continuous star formation in the rings over the past $\sim 300$~Myr. Due to the uncertainties in dating the clusters, we cannot rule out episodic,
$\sim 20$~Myr-long bursts of star formation, 
but the presence of UV-bright
rings in about 10\% of spiral galaxies argues against this possibility.
Although most of the observed SEDs are well fit by a range of models,
some  of the brightest young clusters have excess emission in the IR
 that is not predicted by the models and may be
thermal re-radiation by circumstellar dust. 
The cluster mass functions follow a power-law
distribution with index $-2$,  similar to that
recently derived for the starburst 
in the  merging Antennae galaxies, and extending to $\sim 10^5 M_{\odot}$.
  The lack
of a mass scale means that subsequent evolution of the mass
function is required, if some of the SSCs are to evolve into globular clusters.
The clusters are spatially unresolved or marginally 
resolved, corresponding to $V$-band 
Gaussian radii of less than  a few pc, at 
an assumed distance of 10~Mpc. In NGC~5248,
we report a previously-unknown, 60-pc-radius, inner emission-line ring,
and in NGC~1512, a peculiar compact ($0.1''$ diameter)
 source with an \Hans+[N~II] equivalent
width  of $\sim 7000$~\AA, which may be a so-called Balmer-dominated
supernova remnant.  
\end{abstract}


\section{Introduction}

Galaxy imaging with the {\it Hubble Space Telescope} ({\it HST})
has revealed a population of compact, young star
clusters in a variety of starburst environments 
(Holtzman et al.\ 1992; Benedict et al. 1993;
Whitmore et al.\ 1993; Conti \& Vacca 1994; Hunter, O'Connell,
\& Gallagher 1994; O'Connell, Gallagher, \& Hunter 1994; O'Connell et al. 1995;
Whitmore \&
Schweizer 1995; Meurer et al. 1995; 
Barth et al.\
1995; Maoz et al.\ 1996; Holtzman et al. 1996; Miller et al. 1997; 
Stiavelli et al. 1998; Carlson
et al. 1998, 1999; Whitmore et al. 1999; Alonso-Herrero et al. 2000; 
Drissen et al. 2000; Gallagher, Homeier, \& Conselice 2000a; Gallagher et al. 2000b; Hunter et al. 2000). 
These include
cooling-flow galaxies, interacting/merging galaxies,  
``amorphous'' peculiar galaxies, and galaxies with circumnuclear star-forming
rings. Although such  clusters were known previously from ground-based
and infrared observations (e.g., Arp \& Sandage
1985; Tacconi-Garman, Sternberg, \& Eckart 1996), 
\hst has vastly increased the number of known objects
and the wavelength range and distance to which they can
be studied. The \hst resolution has also allowed 
measurements demonstrating the compactness of the clusters.

The more luminous  among the clusters  ($M_V <
-10$ mag [$L_V > 1.3\times 10^6 L_{V \odot}$], 
according to some conventions, though the limit is somewhat arbitrary
and wavelength dependent) are sometimes termed
 ``super star clusters'' (SSCs). The small radii, high
luminosities, and presumably high masses of these clusters have led
to suggestions that they may remain as bound systems and therefore
could be  young globular clusters.
This view has been reinforced by velocity-dispersion measurements
of several bright SSCs (Ho \& Filippenko 1996a,b; Smith \& Gallagher 2000).
Sternberg (1998) has questioned the ultimate globular cluster fate
of
some of these SSCs. More general debates of the relation between SSCs and 
globular clusters based on their statistical properties can be found
in van den Bergh
(1995, 2000), Meurer (1995),  
Elmegreen \& Efremov (1997), Zhang \& Fall (1999), and Kroupa, Aarseth, \& Hurley (2001).
Recently, new views of SSCs are emerging from infrared (IR) and radio observations
of dense regions in starburst galaxies that may be very young SSCs,
still embedded in their molecular clouds  (Smith et al. 1999; Kobulnicky \& Johnson 1999;
Turner et al. 1999, 2000; Beck, Turner, \& Kovo 2000).
Learning more about SSCs can shed light on
rapid star formation, globular-cluster formation, and the
possible connections between them.

Circumnuclear star-forming
rings in barred spirals comprise possibly the most numerous
class of nearby starburst regions, and offer us an opportunity to
study starburst properties at a level of detail that would be
impossible for more distant or more heavily obscured systems.
The large-scale morphologies and properties of many
circumnuclear rings
have been studied extensively from the ground (e.g., Hummel et al.\
1987; P\'erez-Ram\'\i rez et al. 2000), but
the small sizes of the rings (typically $\sim10$\arcsec\ for nearby galaxies)
require the capabilities of {\it HST\/} for studies of their detailed
structure. Images with WFPC,
WFPC2, and FOC  have revealed large populations of SSCs
 in several circumnuclear rings in nearby galaxies (Benedict et al. 1993;
Barth et al.\
1995; Meurer et al. 1995; Maoz et al.\ 1996; Buta, Crocker, \& Byrd 1999).  
These clusters, barely resolved by
{\it HST}, have effective radii of only 2--4 pc, and luminosities that
range as high as $M_V =$ --14 to --15 mag.  As in other environments,
if such luminous clusters survive as bound entities for $10^{10}$ years, 
they will have
magnitudes comparable to those of old Galactic globular clusters
today.  Circumnuclear rings represent the only known environment in
morphologically normal galaxies in which SSCs are found
in abundance.
  Furthermore, analysis of ultraviolet (UV) 
{\it HST\/} images of five such circumnuclear rings indicates that much
 of the massive star formation in the starburst regions must
occur within the compact clusters (Maoz et al.\ 1996).

Measuring the masses and ages of
SSCs is an important step toward understanding their
future evolution and determining whether they are similar
to young globular clusters.  
 Earlier {\it HST\/} studies of star
clusters in circumnuclear rings (Benedict et al.\ 1993; Barth et al.\ 1995)
were based mostly on single-filter images, and the resulting
interpretation of the cluster properties was uncertain.
Maoz et al. (1996) used an {\it HST} optical image of the circumnuclear ring
in NGC 2997, along with a UV (2300 \AA) exposure, to show that with
two bands (i.e., a UV-optical color), meaningful
 lower limits on the masses and upper limits on the extinctions and
ages of the clusters could be set. 

Recently, Buta et al. (2000) presented
\hst WFPC2 images in five broad UV and optical bands and in \Ha of the 
circumnuclear ring in NGC~1326, another nearby  barred early-type spiral 
galaxy. Comparing their data for the numerous individual detected clusters
to spectral synthesis models, they derived a continuum of ages from
5 Myr to 200 Myr, but with 80\% to 90\% of the clusters being younger than 50 Myr, 
and typically moderate visual extinctions of 0.6-0.9 mag.
The derived masses of the clusters go up to $5\times10^5 M_{\odot}$.
They also noted that many of the H~II regions were devoid of continuum sources,
and concluded that the embedded clusters must have high extinctions, $A_V>2.5$ mag.
 
NGC 1512 and NGC 5248 (Hubble types SBab and SABbc, respectively)
are two circumnuclear ring galaxies
that were included in the Maoz et al. (1996) FOC UV (2300 \AA) imaging
survey. Distances to these nearby galaxies (heliocentric
velocities $v_h = 911$ and 1156 km s$^{-1}$, respectively) are not
accurately known, but are of order 10 Mpc. Maoz et al. detected 43
SSCs in NGC 1512 and 46 in NGC 5248, with similar apparent brightness
distributions. Here we describe subsequent broad-band (approximately
$U$, $V$, $I$, and $H$) and narrow-band (H$\alpha$+[N~II] and Pa$\alpha$) \hst 
images of these galaxies with WFPC2 and NICMOS. 
These observations probe for the first time the spectral energy 
distribution (SED) of SSCs over such a wide wavelength interval.
Apart from providing stronger constraints than before on the
physical parameters of the clusters, they have the 
potential to reveal a large population of clusters
that is obscured by dust. Such an obscured population may
be younger than that which is detectable at shorter wavelengths,
which select only the clusters that have blown off, or emerged from,
their birth clouds. The hydrogen emission-line observations
provide a direct probe of the ionizing flux from massive stars,
and the H$\alpha$/Pa$\alpha$ ratio is a useful reddening diagnostic.
 These observations can therefore
give a more complete picture of SSC properties
and evolution than previously possible.
Both galaxies are at high Galactic latitude and hence undergo little
Galactic reddening ($E[B-V]=0.02$ mag for NGC~1512 and 0.01 mag for NGC~5248;
Schlegel, Finkbeiner, \& Davis 1998); we will neglect these small Galactic 
extinction corrections. 

In \S 2, below, we describe the observations and data reduction.
In \S 3, we analyze the reduced images. We discuss our results
and summarize the main conclusions in \S4.

\section{Observations and Reduction}

WFPC2 images were taken through the F336W, F547M, F814W, and F658N filters
(which correspond loosely to $U$, $V$, $I$, and 
H$\alpha$+[N~II] $\lambda\lambda6548, 6583$). The center of each galaxy 
was positioned on the PC CCD. The 
 rings, which have diameters of $15''$ and $11''$ in NGC~1512 and NGC~5248,
respectively, fit well on the $34''\times 34''$ PC field of view.
Integrations were split into sub-exposures to allow cosmic-ray event rejection.
The dates and total exposure times 
are listed
in Table 1. The individual images were processed by the standard 
Space Telescope Science Institute (STScI) pipeline
(Biretta et al. 1996), which included bias subtraction and flat-fielding.
We combined the sub-exposures using the STSDAS CRREJ task in 
 IRAF\footnote{IRAF (Image Reduction and Analysis
Facility) is distributed by the National Optical Astronomy Observatories,
which are operated by AURA, Inc., under cooperative agreement with the
National Science Foundation.}  to correct
for cosmic-ray hits.

NICMOS Camera-2 images were taken through the F160W, F187W, and F187N
filters.  The F160W filter 
has an effective wavelength of 1.6 $\mu$m, 
at the minimum of the background due
to zodiacal light and thermal emission from the telescope.
 The \Pa line from these
two galaxies  falls well within
the 1\% width of the F187N filter. 
F187W is a broad-band filter centered at 1.87 $\mu$m and was used to provide
a measurement of the continuum underlying \Pans.
The Camera-2 field of view is $19''\times 19''$, also accommodating the
full diameter of the rings.
  At the wavelengths of the bandpasses we used, the Camera-2 resolution
approximately matches the telescope diffraction limit of $0.14''$, 
corresponding to 7 pc at a distance of 10 Mpc. 

For each galaxy-filter combination, the NICMOS observations were split
into four equal exposures, between which the telescope was dithered
in a ``square'' pattern with a step size of $0.5''$, to allow
subsequent elimination of the effects of bad pixels in the detector.
Each individual exposure was read with 18 non-destructive
readouts in MULTIACCUM mode with the STEP64
temporal sampling sequence. This sequence involves logarithmically growing 
readout intervals up to 64~s, followed by evenly spaced intervals
of 64~s for the remainder of the exposure. The MULTIACCUM mode  produces
a large dynamic range and the ability to reject cosmic-ray events.
The NICMOS images were also reduced by the STScI pipeline
 processing system, which
includes bias-level subtraction, nonlinearity correction, dark-image
subtraction, flat-fielding, cosmic-ray rejection, and combination of the 
individual dithered exposures. The pipeline-reduced narrow-band F187N images
of NGC~1512 have a low count rate and show the ``pedestal'' feature common
in NICMOS images, which results from improper bias-level subtraction. We
re-reduced these images with STSDAS in IRAF, 
including the BIASEQ and PEDSKY tasks to treat the pedestal feature.
Although this improved the final results, some remaining background offset 
in one of the four NICMOS quadrants was removed by adding a constant to the
counts in that quadrant.

Photometric calibrations for each camera and filter,
 as listed in Table 1,
were used to convert count rates to physical flux units. The inverse 
sensitivities for the broad and medium-band filters are the latest
circulated by STScI, for a source that has a constant spectrum
in $f_{\lambda}$. The inverse sensitivities of the two narrow-band
filters were obtained by multiplying the inverse sensitivities for a 
continuum source in these filters
by their 
effective bandwidths. The effective bandwidths were calculated by
integrating over wavelength the filter/system throughput curves and dividing
by the throughput at the redshifted wavelengths of the lines for each
galaxy.  For NGC~5248, \Pa
is on the falling part of the filter/system throughput curve, 
where the transmission is 15\% to 35\% lower than at peak, for line-emitting
gas with velocity of $\pm 100$ km s$^{-1}$ in the galaxy's frame.
As a result, its inverse sensitivity value is higher than that of NGC~1512,
and is uncertain, for a given line-emitting region, by $\pm 15\%$.
The effective bandwidths are 39.5~\AA~ for F658N,
213~\AA~ for F187N in NGC~1512, and 275~\AA\ for F187N in NGC~5248. 
We note that, in general, the effective bandwidth is significantly larger
than the full width at half maximum (FWHM), or other filter widths
appearing in the image headers and in STScI documentation. 

In our analysis, we also use the \hst FOC $f$/96 F220W images of these galaxies
presented and analyzed by Maoz et al. (1996). F220W is a broad-band
filter centered at 2300~\AA. See Maoz et al. (1996) for full details
of those data.

To produce continuum-subtracted P$\alpha$ images, we ``blinked'' the
F187N and F187W NICMOS exposures and verified that they were well aligned
(i.e., the telescope pointing did not change between filters, to within 
the resolution limits). For each galaxy, we then scaled the F187W image by the ratio
of the inverse sensitivities of the F187N and F187W filters, and subtracted
it from the F187N image. The \Pa emission in these galaxies comes from
relatively extended regions having weak continuum emission (see below).
The central wavelengths of the two filters are the same, and the continuum
makes only a small contribution to the F187N bandpass in these regions. 
Consequently, we find that the difference images 
are insensitive to the exact method of continuum subtraction, and there is
no need to perform higher-order corrections (e.g., to account for the 
slope of the continuum by utilizing the F160W images).

We used combinations of scaled versions of the F547M and F814W WFPC2
images to produce continuum images at 6580~\AA\ for each galaxy,
for subtraction from the F658N images. The scaling and relative weights 
were calculated from the central wavelengths, the inverse sensitivities,
and the relative exposure times in the three filters, and we found that this
gives a good continuum subtraction, without artifacts such as  conspicuous
negative-flux regions. As with the \Pa images, the good subtraction 
is a result of the mutual avoidance of stellar and emission-line light
in these galaxies.

Quillen \& Yukita (2000) recently
presented an analysis of some of these data.
Their work concentrates only on the emission-line data for
just one of the galaxies discussed here (NGC~1512), and on the
relations between reddening, gas density, and metallicity. Instead 
of the effective badwidths, they 
apparently used the plateau width for the F658N filter
and the FWHM for the F187N filter
 in calibrating the 
narrow-band images of NGC~1512. Their \Ha fluxes are therefore underestimated
by 40\% and their \Pa fluxes by 11\%.
In their reduction, they used the F160W image for \Pa continuum
subtraction. Apart from the relatively large wavelength difference,
which may compromise the accuracy of their continuum subtraction,
the mismatch between the point-spread functions (PSFs) in the F160W
and F187N bands is apparent from their figure.

To enable tracking the SED of individual sources
from the UV to the IR, and for constructing ratios of images in different bands,
we performed a careful alignment of the images. All images of a galaxy
were registered with the F547M image. This was done by ``blinking'' a
given image and the F547M image, and marking the coordinates of the brightest isolated clusters
appearing in the two bands. This proved to be fairly challenging for
bands far from the F547M band, due to the diverse colors of the
clusters (e.g., some clusters that are bright in $V$ are barely visible at $1.6~\mu$m,
and difficult to identify among the myriad of IR sources that 
emerge at longer wavelengths). The cluster coordinates were used to calculate a
transformation between images using the GEOMAP task in IRAF, 
allowing for rotation, translation, and scaling, and the GEOTRAN task was used
to apply the transformations to the images. The emission knots in the \Pa 
image are extended and, as mentioned above, not coincident with the clusters
seen in the continuum bands. However, since all the NICMOS exposures of each galaxy
were already well aligned (including the F187N exposure, which includes
continuum emission), the F160W transformation was applied to the continuum-subtracted
\Pa image. Blinking between the optical and the  transformed IR images shows that
the registration is not perfect, due to local distortions in the NICMOS
images. However,
the mismatch is small enough to avoid any ambiguities in identifying a source
in the various bands.

The transformations between the coordinates of the FOC UV images of these galaxies
and the WFPC2 images were similarly calculated, and the UV sources
tabulated by Maoz et al. (1996) matched with their optical counterparts.
To produce an image of the (\Hans+[N~II])/\Pa ratio, we first matched the 
angular resolution of the \Hans+[N~II] image and the \Pa image by
convolving each image with the PSF of the other image's
bandpass. The WFPC2 F658N and NICMOS2 F187N PSFs were determined using 
the Tiny Tim software (Krist \& Hook 1997). We then 
subtracted a median background
level from each continuum-subtracted, smoothed,
 emission-line image, and divided the two.

Figures 1 and 2 show mosaics of the final galaxy images in the different 
observed bands. Figures 3 and 4 show several color-composite views of 
each 
galaxy, combining data from several bands. Figures 5 and 6 illustrate the
(\Hans+[N~II]) and \Pa images with the main line-emission complexes marked.
 Figures 7  and 8
show the \Hans+[N~II] emission as contour plots superimposed on
the ``$U$-band'' F336W images. Figure 9 gives
the (\Hans+[N~II])/\Pa ratio map for NGC~1512. The large width of the ring
in NGC~5248 (see below) results in a low emission-line surface brightness 
and a noisy (\Hans+[N~II])/\Pa ratio map of limited utility, 
which we therefore do not show.

\section{Analysis}
\subsection{Ring Morphology}

Although similar in most respects (see below), the overall morphologies of the 
circumnuclear rings differ in NGC~1512 and NGC~5248. NGC~1512 has a highly ordered
and narrow ring, with a radius-to-width ratio of $\sim 10$. 
Nearly all of the line emission and the compact sources are concentrated
in this ring.
In NGC~5248, the radius-to-width
ratio is $\sim 2$, and the H~II regions are distributed over a much 
larger area. These differences may be related to the fact that
NGC~1512 is a more strongly barred galaxy than NGC~1512.
NGC~5248 also has a second, inner, ring of $2.5''$-diameter major axis
(120~pc, assuming a distance of 10~Mpc), 
which  is seen most clearly in its \Hans+[N~II] image. To our knowledge, this feature
has not been noted as such before, but it coincides with the inner part of the
spiral found in continuum images by Laine et al. (1998).
 
Comparison of the images of each galaxy in the various filters reveals
a number of features. The appearance of the galaxies changes dramatically
when going from the UV to the IR, because of the growing prominence
of the red populations of the central galactic regions. 
At first glance, the appearance
of the circumnuclear 
rings does not seem to change so drastically between bands, with the
same major complexes of star formation being traceable from the
UV to the IR. Closer examination shows, however, that the structure within 
individual complexes is strongly wavelength dependent. Specifically, the
individual stars and clusters have a large range in color, making the
most prominent sources in one band inconspicuous or absent in another
band. A good example of this is the close pair of bright clusters in the
southeast part of the NGC~1512 rings (top of panel in Figure 1.)
The lower-left member of the pair is considerably fainter in the UV bands,
but is brighter than
the upper-right cluster at 1.6 $\mu$m (see also \S 3.3.5). 
Similarly, a progressively larger
number of sources become discernible as one goes to the IR, and their IR
fluxes are comparable to the IR fluxes of the bright UV sources. This 
large diversity in color and in the bands in which specific sources are
above the detection threshold create the ``Christmas tree'' appearance 
of the color composites in Figures 3 and 4. This phenomenon is treated
more quantitatively and physically in \S 3.3.

The UV images of the galaxies show that the 
brightest regions follow a definite
spiral, rather than ring, structure. The optical and IR images prove, however,
that this effect is produced by spiral dust structures that cross
the rings. As dust obscuration 
becomes less pronounced
at IR wavelengths, the ``gaps'' between the ends of the spirals nearly
close. The true ring morphology of the underlying ionizing population is
seen clearly in the \Hans+[N~II] and \Pa images, in which the spiral structure is no 
longer evident. 

\subsection{ Continuum and Line Emission, and Extinction}
The continuum and emission-line images differ in appearance not only in terms 
of the spiral structure, but in other respects as well.
While the brightest continuum emission from the circumnuclear rings
comes from numerous compact (see below) clusters and stars, the \Hans+[N~II]
and \Pa emission occurs in more extended, resolved clumps encompassing
entire complexes of clusters. In several regions, the line-emitting gas
may actually have a bubble morphology, surrounding a group of young clusters.
In the parts of the rings
showing both strong continuum and line emission (e.g., the bright southern
complex in NGC~1512 -- top of Figure 7), closer examination of the 
\Hans+[N~II] contour map overlaid on the $U$-band continuum image shows that the 
brightest emission-line clumps are at positions that are offset from
the brightest continuum sources. This suggests that expansion due to
winds or ionization
fronts around the clusters clears out the gas and the dust.
Such shell-like or bubble-like morphologies around young clusters are
also seen, for example, in NGC 1569 (Hunter et al. 2000), NGC 5253 (Strickland 
\& Stevens 1999), and 
     NGC 4214 (Ma\'iz-Apell\'aniz et al. 1999).

More remarkably, perhaps (and as already noted above), in some regions
of the rings the line emission and continuum emission are mutually exclusive.
This is conspicuous, for example, in NGC~5248
(Figure 8), where clumps of \Hans+[N~II] emission trace ionizing sources
that are not detected in any of the continuum bands.
This type of emission-line/continuum
avoidance has been noted before in other starbursts (e.g., Ravindranath
\& Prabhu 1998; Buta et al. 2000). It can be seen clearly, for instance, in the \hst composite 
\Ha and continuum image of the merging galaxy pair NGC~4038/4039 
(Fig. 4 in Whitmore et al. 1999). 

To investigate this phenomenon quantitatively, we have measured the emission-line
and continuum fluxes from the main H~II complexes along the rings.
Table 2 lists the integrated \Hans+[N~II], \Pans, and $1.6~\mu$m fluxes, and their ratios,
as measured in circular 
apertures roughly centered on the main emission-line complexes.
The labels and apertures for each complex are marked in Figures 5 and 6. 
The radii of the apertures, also listed in Table 2, were
approximately matched to the size of each emission complex. 
Because of the spatial correlation
between the H~II regions, the definition of what constitutes an individual
region is not unique, and the choices of regions and radii are somewhat 
subjective. These measurements nevertheless provide some rough estimates
of the physical properties of the gas.
Quillen \& Yukita (2000) have compiled a similar table for NGC~1512.
Apart from the incorrect calibrations, mentioned above, there seems
to be an error in the offset coordinates of some of the H~II regions they list
(e.g., the bright complexes 4, 5, 7, and 8 are not on their list). 

Assuming approximate distances of 10 Mpc to the galaxies, the typical diameters
of the brightest H~II regions are in the range 20-50 pc, and the observed  \Hans+[N~II] luminosities
are $10^{37}$ to $10^{38}$ erg s$^{-1}$. As we will see below, based on the
\Pans/\Ha intensity ratios,
the typical  extinction of \Ha is 1--2 mag, or more, if the gas and dust are
mixed, as seems to be the case. The extinction-corrected  \Hans+[N~II] luminosities 
are therefore about an order of magnitude higher than the above numbers.
 For comparison, the \Ha luminosity of the Orion nebula 
is  $10^{37}$ erg s$^{-1}$, and that of 30 Doradus is about
$3\times10^{39}$ erg s$^{-1}$, the rough lower 
limit denoting ``giant H~II regions''
(Kennicutt \& Chu 1988).
 The total \Hans+[N~II] luminosities of the rings, obtained by
integrating the fluxes over the entire rings, are $6\times 10^{39}$ erg s$^{-1}$ (NGC~1512) 
and $9\times 10^{39}$ erg s$^{-1}$ (NGC~5248). Again, the extinction-corrected luminosities
are likely an order of magnitude larger, or about $10^{41}$ erg s$^{-1}$. This is very similar
to the \Ha luminosity derived by Buta et al. (2000) for the starburst ring in NGC~1326. It
 implies an ionizing photon luminosity of $7\times 10^{52}$ s$^{-1}$
and a star-formation rate of about $1 M_{\odot}$ yr$^{-1}$, assuming a 
continuous starburst with a Salpeter initial mass function (IMF)
from 0.1 to 100 $M_{\odot}$ (Kennicutt 1998).

The
 \Pans/$f_{\lambda}$(1.6 $\mu$m)
 ratio, or the equivalent width of \Pans, 
is a good age indicator, since it basically  measures the ratio
of O stars to K and M giants and supergiants, and is insensitive to
reddening because both measurements are at similar wavelengths.
Using spectral synthesis models (see \S 3.3.3 for details) 
for a starburst with a star-formation rate that exponentially decays 
on a timescale 
of 1 Myr, we find that the   \Pans/$f_{\lambda}$(1.6 $\mu$m) ratio has  
a value of $\sim 400~$\AA\ for the first 
few $10^6$ yrs. It then
 declines by 2 orders of magnitude and
becomes of order 1 \AA\ by $\sim 10^7$ years. The nebular continuum
emitted by the gas is included in this calculation.

The observed ratios
in Table 2 are actually lower limits to the intrinsic ratios for
the starbursts producing the lines, because our integration includes
continuum emission from unrelated foreground and background populations
of stars.
 From the observed numbers, which are roughly in the range 50--500~\AA,
 we see that the bright emission-line
complexes have the ratios expected for a mixture of gas and an ionizing
population that is a few million years old. 

The
(\Hans+[N~II])/\Pa ratios, also listed in Table 2, can give some indication 
of the amount of dust mixed with, or along the line of sight to, the gas.
As explained in more detail below, in the absence of reddening, the expected 
(\Hans+[N~II])/\Pa ratio in these galaxies is about 10. The ratios listed in Table~2
are typically  2 to 5 times lower, and thus indicate some
extinction in most of the
complexes.
The (\Hans+[N~II])/\Pa ratio map of NGC~1512 in Figure 9 reveals additional interesting information
regarding the line emission. The
overlaid contours are the \Pa emission, marking the 
regions where there is significant detected flux in this line, 
above a level of $7.7\times 10^{-16}$ erg s$^{-1}$ cm$^{-2}$ arcsec$^{-2}$.
The inner contour corresponds to 
$20\times 10^{-16}$ erg s$^{-1}$ cm$^{-2}$ arcsec$^{-2}$.
 The \Pa image 
is shallower than the \Hans+[N~II] image, because of the lower flux in \Pans, and
the lower sensitivity, higher background, and poorer 
angular resolution in the IR. The \Pa image therefore limits the regions
where the ratio can be measured reliably; outside the outer marked contours,
the ratio is dominated by noise and background residuals.
The whitest regions correspond to a ratio of about 1, while the darkest
areas
have a ratio of $\sim 10$. Concentrating on the pixels within
the contours of significant \Pa emission, we see that most of these have
a greyish shade, which corresponds to ratios of about 4. The same is true
for NGC~5248 (ratio map not shown), but the low surface brightness
of \Pa in that galaxy limits the measurement to a few small regions. 

The dominant error in the ratio is caused by the uncertainty in the background
levels that need to be subtracted from the \Hans+[N~II] and \Pa images prior
to their division. To estimate this error, we created ratio images using 
extreme plausible values for the backgrounds. We find that, within the marked 
\Pa contours, the (\Hans+[N~II])/\Pa uncertainty is $\sim\pm 20\%$, decreasing to $\pm 5\%$
in the regions of highest line emission. This error is compounded by the $\pm15\%$
\Pa calibration uncertainty in NGC~5248 due to the unknown line-of-sight velocity of each clump.

The expected Case B recombination value  of \Hans/\Pans is about $8.5$. The heliocentric
radial velocities of these galaxies (911 km s$^{-1}$ for NGC~1512,
and 1156 km s$^{-1}$ for NGC~5248) places both  lines of the
[N~II] $\lambda\lambda 6548, 6583$ doublet also within the F658N bandpass.
For the nucleus of NGC~1512, [N~II] $\lambda 6548$ is on the blue edge 
of the bandpass, at about 10\% of the peak transmission, while
[N~II] $\lambda 6583$ falls on the flat transmission plateau, along with \Hans.
For NGC~5248, the [N~II] lines are on the rising and falling parts
of the bandpass, each at about 50\% of the peak transmission, and \Ha 
is on the plateau. However, the individual H~II regions in the rotating 
rings can have line-of-sight 
velocities of order 100 km s$^{-1}$ relative to that of the nucleus (e.g., Maoz et al. 1996), shifting 
the lines by $\sim \pm 2$ \AA, and changing somewhat the relative contributions
of each of the [N~II] lines to the counts in the bandpass. 
The nuclear spectrum of NGC~5248 presented in Ho, Filippenko, \&
Sargent (1995) shows the [N~II] lines to be of comparable strength
to \Hans. However, in the circumnuclear H~II regions, the
[N~II] lines, relative to \Hans, are usually much weaker (e.g., Ho, Filippenko,
\& Sargent 1997). Specifically, we have measured the relative line strengths
of [N~II] and \Ha in two of the ring H~II regions that fall along the slit
in the Ho et al. (1995) spectrum of NGC~5248. Assuming these two H~II regions
are representative of the rest, and accounting for the F658N 
transmission at the systemic velocity, [N~II] contributes 14\% to 18\% of  
the counts in the bandpass. The maximum [N~II] contamination, $\sim$23\% to 28\%,
will be for  a velocity of $-100$ km s$^{-1}$ relative to systemic.
We do not have a spectrum of NGC~1512, but assuming it is similar to NGC~5248,
at the systemic velocity [N~II] contributes 22\% to 28\% to the counts, and less
than this for velocities of order  $\pm 100$ km s$^{-1}$. From all these
considerations, in the absence of reddening, we expect the observed
(\Hans+[N~II])/\Pa flux ratio in both galaxies to be in the range of 10 to 12,
substantially above the \Hans/\Pa ratio.  

The typical values integrated over complexes (Table 2) and
observed in the ratio maps, in the range of 2--5,
again suggest that the line-emitting gas
suffers from some dust extinction.
Assuming the
Galactic extinction curve of  Cardelli, Clayton, \& Mathis (1989), with
$A_V=3.1E(B-V)$, a moderate amount of extinction, $A_V\,\approx$ 1.5--3 mag,
 by a foreground dust screen would produce the observed values.
Alternatively, the gas and
dust may be mixed. The observed line ratio is then just the intrinsic 
ratio times the ratio of the optical depths. For an infinite gas+dust slab
($\tau>>1$) the expected line ratio is about 2, similar to what is seen
in many of the regions. It is likely that a combination of these 
two geometries is operating in practice. This would explain the
absence of continuum sources associated with some of the main
emission-line complexes.

Examining in closer detail the ratio maps, they do reveal some small,
dark-coded clumps, often near the peaks of line emission, corresponding
to higher  (\Hans+[N~II])/\Pa ratios, of about 6 to 9. These must be regions
where the gas either undergoes less extinction, or
has a lower ionization spectrum, with a higher [N~II]/\Ha
ratio, more similar to that of the nucleus. This can arise through
excitation by shocks, or photoionization by unusually hot stars, which produce
a hard ionizing continuum (Shields \& Kennicutt 1995; Ho et al. 1997). 
Similarly,
some regions of the rings
also have lower  (\Hans+[N~II])/\Pa ratios. For example, patch No. 6  in the 
west side of the ring in NGC~1512
has a fairly uniform value of  (\Hans+[N~II])/\Pans\,$\approx$ 1--2. The small 
ratio in this patch can also be appreciated by examining directly the 
\Hans+[N~II] and \Pa images in Figure 5.  In NGC~1512,
we do not see any regions that have detected \Pa emission, 
but little or no \Hans+[N~II]. NGC~5248, on the other
hand, does have some \Pa clumps that are faint or undetected in \Hans+[N~II]
(for example clumps No. 15 and 16). The average (\Hans+[N~II])/\Pa in these
clumps is $\ltorder 1$ , with lower values in particular regions of a clump.
This indicates foreground dust, equivalent to the effect
of $A_V\,\approx$ 3--4 mag of Galactic dust,
along the line of sight to some of the line-emitting gas in this galaxy.

We find additional evidence for substantial dust extinction by comparing 
the rings' total ionizing photon luminosities of $7\times 10^{52}$ s$^{-1}$, derived above based on the \Ha flux,
to their UV luminosities. Maoz et al. (1996) measured total flux densities
at 2300 \AA\ for each of the rings, of $f_{\lambda}=8.5\times 10^{-15}$ erg s$^{-1}$ cm$^{-2}$ \AA$^{-1}$, 
or specific luminosities $L_{2300}=10^{38}$  erg s$^{-1}$ \AA$^{-1}$ at the assumed 
10~Mpc distances. They found that $\sim$40\% of this total UV flux
comes from compact sources, and mostly from the few brightest ones.
As we will see below (\S 3.3.4), the SEDs of these bright
clusters are like those expected from clusters that are 1--10~Myr
old undergoing typical attenuations at 2300 \AA\ by factors of 
4 (in NGC~1512)  to 18 (in NGC~5248).
The extinction-corrected luminosities are therefore 
$L_{2300}=4\times 10^{38}$  erg s$^{-1}$ \AA$^{-1}$ (NGC~1512) and 
$L_{2300}=2\times 10^{39}$  erg s$^{-1}$ \AA$^{-1}$  (NGC~5248).
 
 From our spectral synthesis models, we find that a 1~Myr-old starburst
with the above UV luminosities produces $(6-30)\times10^{52}$ s$^{-1}$ ionizing photons,  while
a 10~Myr-old  model with the same UV luminosity produces fewer ionizing photons by two orders of magnitude. Thus,
under the extreme assumption that all the observed UV light is produced by a stellar population that is
only 1~Myr old (or, equivalently, all the UV light is produced predominantly
by O-type stars), the deduced ionizing photon flux is  
similar to that implied by the extinction-corrected \Ha flux. For 
a more conservative assumption about the spectrum of the 
integrated ionizing population, there is a deficit of ionizing photons
by a factor of 1--100.   
This again indicates that the blue sources we see are only a fraction
of the ionizing sources, while some
 are inside the line-emitting gas, and extinguished
by the dust mixed with it. Given the uncertainties in determining
the emission-line extinction and the precise ages of the ionizing
sources, it is difficult to say what the exact obscured fraction is.

A peculiar, compact, emission-line source is found in the northern side
of the ring in NGC~1512 [PC position (452,251), NICMOS position (204,194),
labeled ``A'' in Fig. 5].
In the \Hans+[N~II] image it is one of the highest surface-brightness objects,
but is very compact and round, with a FWHM
 of $0.1''$, and total \Hans+[N~II]
flux of $2\times 10^{-15}$ erg s$^{-1}$ cm$^{-2}$, or an \Hans+[N~II] luminosity
of $2.4\times 10^{37}$ erg s$^{-1}$. In the \Pa image, it is barely detected,
and we estimate the (\Hans+[N~II] )/\Pa ratio to be $\sim$25. 
In the broad bands, this source is detected, if at all, as
a faint, unresolved object. It is undetected in the UV in the FOC/F220W,
with a limit of $f_{\lambda}
(2300~{\rm \AA})<30\times10^{-19}$ erg s$^{-1}$ cm$^{-2}$ \AA$^{-1}$, but
detected with WFPC2 in F336W, with $f_{\lambda}
(3350~{\rm \AA})=(7\pm 2)\times10^{-19}$ erg s$^{-1}$ cm$^{-2}$ \AA$^{-1}$.
It is only barely detected in the F547M band, with 
$f_{\lambda}
(5490~{\rm \AA})\approx(1.5\pm 1)\times10^{-19}$ erg s$^{-1}$ 
cm$^{-2}$ \AA$^{-1}$, i.e., $V\approx 26$ mag, or $M_V\approx -4$. 
At longer wavelengths, in the WFPC2 F814W and 
NICMOS F187W bands, it brightens again, 
with $f_{\lambda}
(8040~{\rm \AA})=(5\pm 2)\times10^{-19}$ erg s$^{-1}$ cm$^{-2}$ \AA$^{-1}$, and 
$f_{\lambda}
(1.6~\mu{\rm m})=(4.5\pm 1.5)\times10^{-19}$ erg s$^{-1}$ 
cm$^{-2}$ \AA$^{-1}$. Given the large equivalent width of \Hans+[N~II] (about 7000~\AA), 
it is possible  that the counts in the broad bands are affected or dominated by
other emission lines. 

In view of the 
discussion above, if the object has a recombination spectrum, then
 most of the flux in the \Hans+[N~II] band
will be 
[N~II]$\lambda 6583$, with a ratio of [N~II]$\lambda 6583$/\Ha of about 2, 
or more if the line emission in reddened. Alternatively, other types of
line excitation may be at work. 
 The properties of this source are similar to those
of ``Balmer-dominated'' supernova remnants (SNRs) that are seen in the Galaxy
(e.g., Tycho's SNR)
and in the Large Magellanic Cloud (LMC; Smith et al. 1991). 
The optical light from these young
(typically $< 1000 $ yr) SNRs consists 
almost exclusively of \Hans, which is thought to arise from a very fast shock
that excites the line through collisional excitation
of neutral hydrogen atoms and charge transfer with protons. 
 Object A may be the first 
Balmer-dominated SNR found beyond the LMC. Another possibility is that object
A is a supernova event similar to 
SN 1986J (Rupen et al. 1987;
Leibundgut et al. 1991) and  SN 1988Z
(Aretxaga et al. 1999, and references therein). These objects
last a long time, and have late-time optical spectra totally dominated 
by \Ha emission.

\subsection{Compact Sources}
\subsubsection{Source Selection}
A primary goal of our work is to improve our understanding of the
numerous discrete sources -- compact clusters and individual stars --
that appear in the circumnuclear rings, by measuring their SEDs.
As seen in the figures above, the
prominence of individual sources can change dramatically, with
some that are bright in one band becoming faint or undetected in
another. We wish to obtain fluxes, or upper limits on fluxes, in
every band for every source that appears in one or more bands.
 
We therefore created a summed ``UV-optical'' image consisting
of the F220W, F336W, F547M, and F814W images, which were first scaled
to all have similar means. The DAOFIND task was applied to this
combined image to create a list of discrete sources. The complicated, uneven
background and crowding of the sources mandated that the output
of this task be edited manually, to delete false detections and
append obvious misses. Such a procedure was found to be unavoidable
also in previous work on these and other ring galaxies (Barth et al. 1995;
Maoz et al. 1996). We verified that no obvious sources appearing in
the individual optical-UV bands had been overlooked. We attempted to
locate all sources in the regions of the rings, but not in the more exterior
regions, which are outside the fields of view of the FOC and NICMOS, nor
interior to the rings, where the bright stellar background makes it 
progressively more difficult to detect sources at smaller radii and at 
longer wavelengths.  This procedure cannot
be considered objective, but our compilation should be complete for the 
brightest sources in each ring.
In any event, there are few clusters outside or inside
the rings. Altogether, 325 
optical-UV sources were selected in NGC~1512 and 397 in NGC~5248. 

Most of the sources in the NICMOS 1.6 $\mu$m exposures are not obviously
detected in the optical images. We therefore applied the same source detection
procedure separately to the F160W images. Of the 221 IR sources in NGC~1512,
150 were not selected in the UV-optical image. In NGC~5248, 110 of the 191
IR sources were not selected in the UV-optical image. The combined UV-optical-IR
lists of sources thus consist of 475 sources in NGC~1512 and 507 sources in NGC~5248.

\subsubsection{Photometry}
Aperture photometry at the positions of the sources in the final
list was carried out on the images in each individual band
 using the PHOT task in IRAF. For the WFPC2 images, apertures
of radii 1, 2, and 3 pixels were used.
 The 2-pixel-radius apertures are best suited
for WFPC2 PC photometry, since they are large enough to minimize photometry errors
due to pixelation and to centering errors, but small enough to minimize
the effects of crowding on the photometry (e.g., Barth et al. 1995). 
The NICMOS
images have a larger PSF and a coarser pixel scale, with
one NICMOS pixel corresponding to about 1.67 PC pixels. To avoid uncertainties
in the photometry due to the resampling and registration of the NICMOS images,
NICMOS photometry was performed on the original images. The positions
on the NICMOS frames 
of all the UV-through-IR-selected sources were
computed, based on the geometric transformations described above. We used
an aperture of radius 1.79 NICMOS pixels, which is equivalent in angular scale
to 3 WFPC2 PC pixels. This was chosen as a compromise, given the constraints,
on the one hand,
to match approximately the PC apertures in order to measure the same objects
in the optical and IR bands and avoid crowding effects while, on the other hand,
to avoid overly small apertures that will be susceptible to pixelation
and centering errors.

Aperture corrections need to be applied to the measurements to account 
for the fact that not all the light of the  PSF is included
within the aperture. As discussed in \S 3.3.7, the individual sources in these
galaxies appear to be unresolved or marginally resolved.
We therefore calculated the aperture corrections by generating artificial
PSFs for each of the WFPC2 and NICMOS bands using the Tiny Tim software,
and finding the ratio between counts within the radii used for the photometry and the 
radii corresponding to $1.4''$, which should include 99\% of the light 
according to the STScI website. 

Photometry errors due to photon statistics and background uncertainties
were calculated by the PHOT task. To this, we added in quadrature
a calibration uncertainty of 5\%, which includes the effects of
flat-fielding errors, charge-transfer efficiency and 
UV contamination variations in the 
WFPC2 detectors, uncertainties in aperture corrections, and
uncertainties in the absolute sensitivities of each of the bands
 (see the WFPC2 and NICMOS Data Handbooks). 

After aperture correction and photometric calibration, all of the
photometry was consolidated into one list for each galaxy, including the 2300 \AA\ 
fluxes from Maoz et al. (1996). Tables 3 and 4  show the
final photometry ordered by brightness in the F547M filter,
in decreasing order, for each of the galaxies. 
The full tables for all the measured sources in both galaxies 
are available in the electronic version of the journal.

Figure 10 is a color-magnitude diagram showing, for both galaxies,
 all of the sources detected in F547M and F814W. For ease of comparison
with other work, the data in this figure have been converted to 
standard $V$ and $I$ magnitudes according to the transformations of
Holtzman et al. (1995). The $V$ magnitudes also appear in Tables 3 and 4.
The figure shows that the brighest sources are blue, with $V-I\approx 0$.
The sources in NGC~5248 have a larger spread in color
than those in NGC~1512, and are typically 0.5-1 mag redder.

Figures 11--12 plot the 
SEDs of the brightest
individual sources in each galaxy. For clarity, 
only the 25 brightest  
sources in each galaxy are shown,
separated into different panels according
to brightness in $V$, and error bars are omitted.
Missing measurements at 2300 \AA\ and at 1.6  $\mu$m are for objects
outside the field of view of the FOC or NICMOS, or, for the fainter
objects, are below the detection limits.

A cursory examination of the SEDs again shows that almost all
of the bright sources are blue, with spectra that can be approximated
as $f_{\lambda}\propto \lambda^{\beta}$, with slopes $\beta$ of  $-1$ to $-3$. 
These sources apparently undergo little reddening.
 There are no
bright sources that have SEDs which rise to the red.
The most luminous known red supergiants have $M_V\approx -7$~mag,
or, at 10~Mpc,  $V\approx 23$ mag and $f_{\lambda}
(5490~{\rm \AA})\approx2\times10^{-18}$ erg s$^{-1}$ 
cm$^{-2}$~\AA$^{-1}$. While many of the sources in the rings, including
some of the more luminous ones, are only a few times brighter than this,
 their blueness excludes the possibility that they
are individual red supergiants, rather than young clusters.
 As for individual blue supergiants, they can
have even higher luminosities, up to $M_V =-10$ mag, 
but they are extremely short-lived and rare,
and hence are not plausible candidates  for explaining any but a handful
of the observed
sources. If, for the purpose of a rough appraisal of the content
of the clusters, one assumes the UV is dominated by the light of main-sequence
B0 stars, the observed fluxes (uncorrected for extinction)
of the 25 brightest clusters correspond to 5 to 500 such stars.
For reasonable IMFs, the total masses of these 
clusters are of order $10^4$ to $10^6 M_{\odot}$. 
Masses in this range also result from our detailed model fitting (\S 3.3.6).
   
It is only the sources with  $f_{\lambda}
(5490~{\rm \AA})\ltorder 2\times10^{-18}$ erg s$^{-1}$ that may, indeed,
often be individual supergiant stars.  However, some of
these objects may also be reddened SSCs
and old, or low-mass, clusters. For reference, old globular clusters have
$M_V\approx -7$ mag, similar to the most luminous red supergiants. Thus,
some of the faint objects could also be old globulars from a 
previous starburst. This possibility is examined in more detail below.

\subsubsection{Cluster Models}

To interpret the measured SEDs, we use the population and
spectral synthesis code STARS 
(Sternberg 1998; Thornley et al. 2000)
to generate model SEDs of clusters with a range of 
ages and assumed foreground extinctions.
For the population synthesis, STARS employs the Geneva stellar evolutionary
tracks (Schaller et al. 1992). In the computations presented here, we assume
solar metallicity tracks.  
We also assume a time-independent power-law IMF,
$ dn/dm \propto m^{-\alpha}$, between a lower and an upper  mass limit.
For the
spectral synthesis of the cluster SEDs 
we adopt the procedures described in Thornley et al (2000).
Briefly, the spectral energy distributions (SEDs) for individual stars are
taken from a ``hybrid grid'' generated from two libraries of 
theoretical model atmospheres.  For stars with effective temperatures
$< 19000$ K we use the Kurucz (1992)
models which assume atmospheres in local 
thermodynamic equilibrium (LTE). For hotter stars ($25000 - 65000$ K)
we adopt the more realistic 
non-LTE models of Pauldrach et al. (1998).  The SEDs
for intermediate temperatures are obtained by interpolation between
the Kurucz and the Pauldrach et al. models. 
A more detailed discussion and comparison
of the library SEDs is presented in Thornley et al. (2000).  

We have compared some
of our broad-band photometric computations with those of
 other codes, such as STARBURST99 (Leitherer et al. 1999), and
find good agreement. 
For example, our computed $V$ magnitudes agree
well (generally to within 0.3 mag)
with those of Leitherer et al. for both continuous and decaying
bursts and a range of IMF 
 slopes and upper mass cutoffs.

Within the scope of the present study, 
we have limited ourselves to models with
a Salpeter (1955) IMF with upper and lower
mass limits $M_{up}=120  M_{\odot}$ and $M_{low}=1 M_{\odot}$,
solar metallicity, and an exponentially decaying star-formation rate
with a characteristic timescale of 1~Myr.
Solar metallicity is a reasonable
assumption for circumnuclear regions of barred spirals, where
radial mixing occurs (e.g., Roy \& Belley 1993; Martin \& 
Belley 1997), and, in any event,
the broad-band properties
are weakly dependent on metallicity (e.g., Leitherer \& Heckman 1995).
We find that the model calculations are only weakly affected if, 
instead of a Salpeter slope of $-2.35$ we assume an IMF slope
of $-2.5$, or if the burst timescale is 5~Myr instead of 1~Myr.
Figure 13 illustrates the spectral evolution of a model cluster. The five
curves correspond to 
ages of 1, 10, 30, 100, and 300~Myr, with the steepest
SED corresponding to the youngest age, and then fading with age. An arbitrary 
vertical shift has been applied to the models as a whole (but not relative
to each other).  A vertical shift is equivalent to a change in the mass
of a cluster or in the distance to a galaxy.

The light from a 1~Myr-old model cluster is dominated at all bands by
main-sequence O and B stars, and by evolved massive stars. The overall spectral
slope is approximately a power law with slope $\beta \approx -3.5$.
By 10~Myr, the SED develops a change of slope longward of
 the Balmer edge, where the light becomes dominated by red supergiants,
but the slope at shorter wavelengths also flattens, as
the most massive stars disappear. The continued evolution of the cluster
is characterized mainly by  a slow overall fading, along with a strengthening
of the Balmer jump, and only mild changes of slope. Note that there 
is less than a factor of 10 difference in $V$ luminosity between the
oldest and youngest model clusters we have considered.
Therefore, coexisting clusters
within this wide range of ages can be detected, in principle. As we  
show below, however, the steep mass and luminosity functions of the clusters create
an age bias that will cause young clusters to appear highly over-represented
in a given sample.

Figure 14 illustrates the effects of extinction by foreground dust
on one of the models, and was calculated using
the Galactic extinction curve of Cardelli et al. (1989), with
$A_V=3.1E(B-V)$. Although many other extinction laws are possible, we will,
again, limit our investigation to this one.

\subsubsection{Color-Color Diagrams}

One way of comparing the models and the observations is by means
of color-color diagrams, as has been done, for example, 
 by Whitmore et al. (1999) and Buta et al. (2000).
A problem with this method is that, for particular choices of bands,
the age trajectories of model clusters sometimes ``backtrack'' on themselves,
making it difficult to uniquely determine the ages of some of the 
observed clusters. Furthermore, age and extinction may move points in the 
diagram in a similar direction, so that there is a degeneracy in these
two parameters. We have found that most of the color-color plots we
can make with our available  broad-band data suffer from the same problems.
However, the use of the UV (2300~\AA) band can partly solve this.
Figure 15 shows the models and the data on a plot of $2.5 \log [f(2300~ {\rm \AA})/f(3350~ {\rm \AA})]$
vs. $2.5 \log [f(3350~ {\rm \AA})/f(5490~ {\rm \AA})]$ (essentially, $U-UV$
and $V-U$, up to additive constants). The model fluxes were calculated by integrating the model SEDs
over a bandwidth equal to the filters' FWHM.
 Cluster aging trajectories for several values of foreground
 extinction
are shown. Ages, in Myr, are marked in large numerals along the top
trajectory. One sees that, in this color-color space, 
the individual trajectories are well stretched out, partly 
solving the ``backtrack''
problem. The reddening shifts the curves in a direction nearly orthogonal 
to the aging direction, partly alleviating
 the age-extinction degeneracy. Unfortunately,
it is possible to place on this plot only the $\sim 40$ clusters in each galaxy that were
detected in the relatively shallow pre-refurbishment FOC UV images. 
Nevertheless, these plots are revealing.

Figure 15 plots with progressively smaller symbols
sources that are fainter in $V$. Typical error bars are shown
in the corner. To avoid cluttering, only the brightest clusters are labeled
according to their designations in Tables 3 and 4.
 It is clear that the brightest clusters are very young,
with ages generally less than 5~Myr for the 15 brightest clusters, and 
generally less than 10~Myr for all the UV-detected clusters. Furthermore,
the extinctions are quite moderate, with the brightest sources clustered
around visual extinctions of $A_V=0.5$ mag in NGC~1512 and $A_V=1$ mag in NGC~5248. 
These correspond to the attenuations at 2300 \AA\ by factors of 4 and 18, 
respectively, already mentioned in \S 3.2.

To illustrate the form of a color-color diagram in another band, for which
most of the sources do have measurements, Figure 16 plots
the same 
model trajectories in the plane of
 $2.5 \log [f(3350~ {\rm \AA})/f(5490~ {\rm \AA})]$
vs. $2.5 \log [f(5490~ {\rm \AA})/f(8040~ {\rm \AA})]$ (essentially, $V-U$
and $I-V$, up to additive constants). For clarity, 
only the models are shown in this plot.  
The problems of backtracking and age-extinction degeneracy are evident.
In Figure 17, the data for
 NGC~1512 are added to a similar plot, which
shows additional extinctions, but omits some age labels and all
cluster labels for clarity. 
The 20 clusters brightest in $V$ are again shown as the largest points, and each set of the next 100 
brightest sources is plotted with progressively
smaller dots. Most of the clusters have separations 
from at least some model trajectories that are consistent with the statistical
and systematic calibration uncertainties of the measurements, as
represented by the typical error bars plotted. As in 
Figure 15, the diagram indicates that most of the bright clusters are
young and only mildly reddened. However there
are two bright sources above, and to the left of, the others, that are
simultaneously very blue in $V-U$, but red in $I-V$, and are
about 0.5 mag from the nearest model. (These are sources Nos. 7 and 17 in
Table 3.) This ``IR-excess'' phenomenon
is examined in more detail below.
For most of the fainter sources, it is
impossible to disentangle age and extinction in this plot, and the
larger measurement errors exacerbate the problem. Some of the faintest
sources are probably individual stars. 

\subsubsection{Model Fitting -- Cluster Ages and Extinctions}
Another method of estimating ages and extinctions, which exploits
all the available broad-band information, is to directly compare
the modeled and observed SEDs. To this end, we have generated
a grid of models with ages between 1 Myr and 300 Myr, and extinctions
between 0 and $A_V=3$ mag. Each model SED was multiplicatively
scaled to best match each observed SED and the $\chi^2$ residual of the fit
calculated. Figure 18 shows examples of three observed SEDs in NGC~5248
and their model fits. For the fainter cluster displayed, 
two models are illustrated --- one that
fits well, and one that is marginally consistent with the data.
This demonstrates the power of this diagnostic technique when
measurements in numerous bands over a wide wavelength range are available.
For most of the observed SEDs 
we can find at least one model that provides an
acceptable fit. Given the observational uncertainties in the SEDs and the
remaining age-extinction degeneracy in some cases (e.g., for some clusters
without a UV measurement), more than one model can fit most of the observed
sources.  

However, some sources are poorly fit by all models we have considered.
Surprisingly, it is the brightest (and bluest) sources in both galaxies
that often suffer from this problem, even though these are least susceptible 
to measurement errors due to, e.g., crowding and background uncertainties.
Figure 19 shows examples of SEDs and
best-fit models for four such clusters in NGC~1512 (Nos. 2, 3, 4, and 8 in
Table 3). Clearly, the 
poorness of the fits (which are all formally rejected)
 is due to the $1.6~\mu$m measurements, which are 2-5 
times higher than the predictions of the models that reproduce well
the data in the other bands.

It is unlikely that this IR deviation 
in some of the bright sources is
a calibration problem, since most of the sources in both galaxies
{\it are} well fit
in all bands, including the IR.
The phenomenon can actually be seen directly by examining,
in  Fig. 1, the two close
bright clusters near the 12 o'clock position. These are clusters
Nos. 1 and 5 in Table 3 (i.e. the brightest, and fifth brightest, in $V$),
and their SEDs appear in Fig. 11. 
Cluster No. 1 is several times
brighter than cluster No. 5 in all bands, except in the IR, where clearly
No. 5 overpowers No. 1. Cluster No. 1 is about as blue as
a cluster can be, so if one changed the IR calibration
to make the SED of No. 5 fit the models, cluster No. 1 (and many
others, too) would be too blue (i.e., too weak in the  IR) to fit any model.
This, again, argues against a calibration problem.
 One could invoke a systematic error,
such as saturation,
that affects only the brighter sources. However, saturation would
make bright sources appear anomalously faint, rather than bright, and, in
any case, no such effect has been reported with NICMOS
 for sources of this brightness.
    
We believe that the observed IR excess is real, and signals some
physical element that is absent from the models. As an example,  
the dashed line in Figure 19 shows the effect of adding to the lower model
a $2000$ K blackbody, scaled to reproduce the IR excess 
of the fainter cluster in the diagram. The excess in this object is 
large enough that it seems to appear both in the $I$ and the $1.6~\mu$m bands,
permitting some constraints on its form.  
If the excess is from dust which is heated by the 
cluster stars to $2000$ K, its distance from the stars is $\ltorder 100$
stellar radii; that is, the dust must be circumstellar.
  
Tables 5 and 6 give the best-fitting model parameters for the 
sources in the two galaxies, with the full tables again provided in
the journal's electronic edition.
Figure 20 shows
 the ages and extinctions for the sources in the two 
galaxies, regardless of the acceptability of the fits. Given the IR-excess
phenomenon described above, the best fits can still provide good estimates
of the ages and the extinctions. Since the parameters of
sources detected in only a few bands are poorly constrained, we fit only 
those sources detected in four or five bands (258 sources
in NGC~1512 and 328 in NGC~5248). Point sizes are proportional to 
$V$ brightness, and points have been given small arbitrary vertical and
horizontal shifts to avoid overlap. The vertical striping pattern in the 
plots is the result of the discrete grid of models used.
There is generally good agreement between the individual 
ages and extinctions
deduced from the 
color-color diagrams in Figure 15 (for those clusters that have a UV 
measurement) and those found from model-fitting, as listed in Tables 5 and 6.
The model-fitting, however,
includes data from two additional bands, and is therefore 
sometimes more constrained.

In NGC~1512, most of the sources are best fit by model clusters that 
are 20~Myr or younger, and have low extinctions, $0<A_V<1$ mag. 
In NGC~5248, the youngest and brightest sources are somewhat redder, with
$A_V \approx 1$ mag,
as already concluded on the basis of color-color diagrams. 
The remaining
sources cover a larger range in age than in NGC~1512, but here too, almost
all have extinctions $0<A_V<1$ mag, and ages 40~Myr or younger. 
There is, however, a significant fraction of clusters in both galaxies
that are best fit with older models, of ages $\sim 100$~Myr.

At first sight, this result is 
contrary to that expected,
 if the starbursts in the rings were continuous.
The clusters fade by a factor less than 10 over 300~Myr (see  Fig. 13),
and the massive ones thus remain well within our dynamic range for this
entire age range. In a continuous burst scenario, one would therefore naively
expect many more clusters to have large ages, and one would conclude
from the  figures
that the starbursts in each of these galaxies are episodic events
which last only several tens of Myrs.    
However, we show here that the scarcity of old clusters may be the result
of an age bias which causes young clusters to be over-represented at 
every luminosity.

Many studies of SSCs in various environments have shown that the
clusters follow a steep luminosity distribution, $N(L) dL \propto L^{\alpha} 
dL $, with $\alpha\approx -2$ (e.g., Meurer et al. 1995; Barth et al. 1995;
Maoz et al. 1996;
Whitmore et al. 1999; Carlson et al. 1999). 
Figure 21 shows the extinction-corrected
$V$-band luminosity functions
for the clusters in NGC~1512 and
NGC~5248, separated into young ($\le 15$~Myr) and old ($> 15$~Myr)
categories. Each cluster's age and extinction are based on its best-fit model.
We have also plotted on the diagrams suitably normalized  
$N(L) dL \propto L^{-2} dL$ power laws. The figures show that the cluster
$V$ luminosities in the two galaxies, when considering only the brightest
clusters that are less susceptible to incompleteness,  
are consistent with  an $L^{-2}$ distribution, as has already
been demonstrated for their 2300 \AA\ luminosities by Maoz et al. (1996). More 
importantly, the figures show that the individual ``young'' and ``old'' 
categories appear to follow the same distributions, although the small
number of old clusters makes this statement tentative. Recalling the 
relatively small, but significant, fading of clusters with age, the evolution of
the young stellar population can be simply described by a horizontal
shift to the left of the luminosity functions shown in logarithmic
coordinates in Figure~21. Because of their steep slopes, such 
effective age correction would make the old clusters intrinsically 
more abundant
than the young ones, even for a luminosity shift by a factor of only a few.
Thus, the fact that one sees relatively so many young clusters in these
galaxies is not necessarily because the entire starburst is young. Instead,
it could be the result of the fact that there are many more low-mass
clusters than high-mass clusters. When the low-mass clusters are young,
the moderate rise to a higher luminosity makes them more abundant than
the older, higher-mass clusters that have the same luminosity. 
    
The relative numbers of young and old clusters are thus consistent with 
continuous star formation. If, on the other hand, the star formation were
episodic, we would expect few or no old clusters to be detected at all.
Turning around the question of a continuous versus an episodic starburst,
we can then ask whether an episodic starburst can be ruled out.
One must recall that the best-fit models are not the {\it only}
models that can fit each observed SED. Even with the considerable leverage
afforded by the large wavelength range of our data, there is still
degeneracy in assigning ages and extinctions to some clusters. Figure 22
shows, for every well-observed SED, the youngest model that is marginally
consistent ($\chi^2 \le 2$ per degree of freedom).
These models are also listed in Tables 5 and 6.
The plots show that, if pushed, the data allow all but a handful of sources
to be modeled as young clusters. It is thus possible that there are actually
few old clusters in the galaxies. This could come about if 
the starbursts in the 
rings occur in $\sim 20$~Myr-long  episodes, with quiescent
intervals of $\sim 100$~Myr, or more. The same phenomenon could also 
occur if most of the clusters do not survive for longer than $\sim 20$~Myr.  

For completeness, Tables 5 and 6 also list, for each well-observed SED,
 the {\it oldest} model that is marginally acceptable. One sees that,
in this case, there are many clusters that cannot be older than a few Myr,
as already concluded based on the color-color diagrams.
 
\subsubsection{Model Fitting -- Cluster Masses}

A total cluster mass can be associated with a model SED with a particular
luminosity normalization. 
 The total mass is that which is reached asymptotically by the 
decaying burst of star formation, 
but without accounting for mass loss from the cluster due to stellar 
winds and supernovae, which is theoretically difficult to predict.
The total mass in an observed cluster
is necessarily uncertain when derived based on photometric
(rather than kinematic) data. Most of the mass is in 
low-mass stars on the main sequence, which contribute negligibly to the 
light at any band. The total mass
would increase by a factor of $2.55$ or decrease by a factor of $\sim 3$
 if the low-mass cutoff were at $0.1 M_{\odot}$ or $5 M_{\odot}$, respectively, instead
of at $1 M_{\odot}$. For young ($\sim 1$ Myr) clusters, the total mass
would increase by about 50\% if the upper-mass cutoff were at  
$30 M_{\odot}$, rather than 
at $120 M_{\odot}$, because in the latter case, the very massive stars briefly
dominate the luminosity, and fewer intermediate-mass stars are needed
to explain the observed flux. Thus, compounding the theoretical uncertainties
with the observational errors (e.g., the luminosity, and hence the
mass, scales with the distance squared, and the distances to these galaxies
are poorly known),
  the total masses we will derive below are uncertain
to at least an order of magnitude, 
although this uncertainty may involve a systematic shift for all clusters,
if all are similar in their star-formation parameters.   

Recently, Zhang \& Fall (1999) have used 
age and extinction estimates to show that, 
in the merging galaxies
NGC 4038/9, the mass function of the clusters follows a power law
similar to the luminosity function, $N(m) dm \propto m^{-2} dm $,
over the range $10^4\le m \le 10^6 M_{\odot}$. The mass function
is important for understanding the relation between SSCs and old
globular clusters. 
Our model fits to the clusters in NGC~1512 and
NGC~5248 assign masses to each cluster, and allow extending this
type of analysis  to two more galaxies and a different starburst
environment. Tables 5 and 6 list the masses for each of the
fitted models. Figures 23 shows the mass functions for the clusters,
based on the best-fit models.
Although we are limited by incompleteness at low masses (low-mass 
clusters are faint) and small numbers at high masses (high-mass clusters
are rare), the data are consistent with the same power-law mass distribution
seen in NGC~4038/9. 

The few, most massive, clusters in NGC~1512 and NGC~5248 have masses of 
$\sim 10^5 M_{\odot}$. Zhang \& Fall (1999) found several tens of clusters
in NGC~4038/9 with masses between $2.5\times10^5 M_{\odot}$ and
$10^6 M_{\odot}$. However, in making this comparison, two points need
attention. First, the models of Zhang \& Fall are generally similar to ours,
but assume an IMF with a low-mass
cutoff of $0.1 M_{\odot}$, and hence their mass estimates 
 are systematically higher
by a factor of 2 than ours. Second, NGC~4038/9 is a more extended starburst
than those in the nuclear rings, and therefore includes more clusters --
Zhang \& Fall derive masses for about 1600 clusters, whereas we do so
for only about 300 in each galaxy. Given the steeply falling mass
function, there is a low probability of finding in the ring galaxies
clusters with the same masses as the few most massive ones in the
merging galaxies. After accounting for these differences, the fraction
of the cluster population above a given mass is similar in all three
starburst systems.

\subsubsection{Cluster Sizes}

While some early studies with \hst of SSCs in starburst environments
claimed that the clusters
were resolved, with sizes of order 10 pc (e.g., Whitmore \& Schweizer 1995), 
recent studies of
galaxies at $\sim 10$ Mpc distances have generally
concluded that the clusters are unresolved, or marginally resolved
with the PC on WFPC2, corresponding to sizes of just a few pc or less. 
Specifically for the two galaxies examined
here, Maoz et al. (1996) examined the radial 
profiles of the clusters in aberrated FOC F220W images. They concluded
that the flux profiles, if modeled as Gaussians convolved with the
PSF, have a dispersion ($\sigma$) near or below the resolution limit, 
corresponding to a Gaussian radius of $\ltorder 2$ pc. We re-examine the 
size question in the present, post-refurbishment, images of the clusters
in NGC~1512 and NGC~5248, now taken in optical and IR bands.

We have carried out two tests to gauge the degree to which the sources
are resolved. First, we studied the azimuthally averaged radial profile
of the brighter sources, as in Maoz et al. (1996), but starting from a radius of 
1 pixel and outwards. We attempted to fit the radial profiles with those of
artificial PSFs produced by Tiny Tim (see \S 2), or with those observed
for some of the brighter isolated sources in the images themselves (to see
if there is a range of source sizes in the data). We find that,
as in the UV, the sources seem to be unresolved or marginally resolved 
in the optical and IR. The result is not surprising in the IR,
since the angular resolution is considerably worse than in the optical.

Second, we compared the ratio of counts within 1-pixel and 3-pixel radii that
 we obtained for the clusters in the
F547M band (which has the best angular resolution) to the ratio obtained
for artificial PSFs. Various authors (e.g., Barth et al. 1995; Whitmore
et al. 1999)
have used this so-called $\Delta m_{1-3}$ test, or similar indicators,
 as a way of measuring sizes,
with $\Delta m_{1-3}$ significantly larger than that of the PSF indicating
resolved sources. Tables 3 and 4 list $\Delta m_{1-3}$ for all sources.
 We find that the Tiny Tim PSF for F547M has $\Delta m_{1-3}$=0.79 mag,
in good agreement with the Barth et al. (1995) value of 0.83 mag. 
To study the effect of the 
PC pixelation on $\Delta m_{1-3}$, we created also a Tiny Tim PSF
in which the position of the source was on the corner of a pixel. (This
is not a standard Tiny Tim option, and required creating a subsampled PSF,
shifting and rebinning it, and convolving it with a kernel that simulates
charge diffusion.) For this PSF, $\Delta m_{1-3}=1.02$ mag.

To get a direct empirical evaluation of $\Delta m_{1-3}$, one would ideally
measure it for obvious foreground stars appearing on the same exposures,
but there are only one or two such stars in these high Galactic latitude
exposures. Instead, we measured $\Delta m_{1-3}$ for the 100 brightest stars
in an F547M image of the globular cluster NGC~6441. The total counts for these
stars are comparable to those of the brightest sources in the two galaxies,
and we used only the less crowded half of the PC1 image of the globular
cluster, where the crowding is similar to that in the circumnuclear rings.
 Figure 24 compares the distribution of $\Delta m_{1-3}$ for the globular
cluster stars with the values for the 10 brightest clusters in each galaxy.
Also shown, with filled symbols, are the two measurements based on ideal
artificial PSFs.
The filled triangle represents
a red, isolated, $V=21$ mag object in NGC~1512 (PC1 coordinates 664, 625), 
possibly a foreground star. A similarly red object (not plotted in Fig. 24)
is found in NGC~5248 (PC1 coordinates 425, 385), 
in a high surface-brightness region, and has $\Delta m_{1-3}=2.12$.
While the distribution of $\Delta m_{1-3}$ for the stars indeed peaks near
0.8, there is a tail to higher values of $\Delta m_{1-3}$, which encompasses
many of the SSCs.  We have repeated this experiment with the F547M image of 
another globular cluster, NGC~5139, which has a less crowded stellar field.
In this case, there appears to be less of a high-$\Delta m_{1-3}$ tail,
but the number of stars is small.
Examination of the cases with $\Delta m_{1-3}>1$ in the first globular
cluster shows that these are stars projected close to other stars of comparable
brightness, confirming that source crowding can have a major influence on
$\Delta m_{1-3}$.

The comparison with the globular clusters is not straightforward. The 
globular clusters were observed about two years before the galaxies, when
the telescope focus may have differed. Their exposures are only 80~s and 50~s,
much shorter than those of the galaxies, and may hence suffer less from
image-degrading effects such as jitter. The stellar crowding in the globular
clusters differs in its details from the crowding among the sources in the
galaxies, which are also superimposed on a complex diffuse background.
Given these uncertainties, 
we conclude that the bright clusters in NGC~1512 and NGC~5248
which have  $\Delta m_{1-3}> 1.1$
are possibly resolved, barring some of the possible systematics we cannot
address. For the remaining bright clusters, it is unclear whether 
the $\Delta m_{1-3}$ values, which are only slightly larger than those of
isolated PSFs, are due to crowding or to truly resolved sizes.

We suspect that various reports
of resolved SSCs in other starburst galaxies (e.g., Carlson et al. 1999) 
based on comparison of $\Delta m_{1-3}$ analogs
to artificial PSFs, neglecting the various systematic 
effects, may have overestimated the SSC sizes.
Some of these other galaxies are significantly
more distant than those studied here (e.g., $\sim 50$~Mpc for 
NGC~3597 in the Carlson et al. 1999 study). At such large
distances, there is  a larger chance of size errors due to crowding
and confusion, especially since the brighter clusters tend to be
spatially correlated.
In the nearest examples, such as NGC~1569, 
the SSCs appear extremely compact as well, suggesting
a distance-related bias. 
Since our tests indicate that the sources are unresolved or
marginally resolved, we will not attempt further to derive physical sizes.
Instead we note that, assuming a generic distance of 10 Mpc, the Gaussian
radii of all the discrete sources are $\ltorder 2$ pc.

\section{Discussion and Conclusions}

We have analyzed broad-band and narrow-band images of two circumnuclear
ring galaxies, obtained with the fine angular resolution of {\it HST}, and
spanning a large continuum wavelength range, from 0.23 to 1.6 $\mu$m,
and two main emission lines, \Ha and \Pans. The data provide a uniquely
detailed look into the properties and interrelations of the gas, the dust, and
the compact clusters.  Our main findings are as follows.

There is only a loose spatial correlation between the observed young 
clusters and the ionized gas in the rings. In some H~II regions, no
continuum sources are seen, and the 
presence of an embedded young population is revealed only by means of the
high \Pa equivalent width. In others, the ionized gas is in shells or
bubbles surrounding complexes of clusters, which are apparently in the
process of destroying their birth clouds. The \Hans/\Pa ratios of the
line-emitting gas generally indicate only moderate extinction by a
foreground screen, or, more realistically, an optically
thick mixture of gas and dust. The latter option would explain the severe
extinction, even at IR wavelengths, of the light from the young stars
embedded in many of the clouds. Ionizing photon budget estimates
also suggest that a large fraction of the ionizing stellar population
is hidden. Unfortunately, because the line emission is probably also 
strongly extinguished by dust, it is difficult to say what this fraction is.

Comparison to other imaging studies of nearby starbursts, whether in 
rings or other configurations (e.g., Whitmore et al. 1999; Buta et al. 2000;
Lancon et al. 2001),
 reveals a recurring pattern, whereby the
continuum and line emission come from distinct sites. Although the
visibility of the continuum sources depends on their ages (it takes time
for them to break out of their clouds), there seems to be a strong
element of geometry, orientation, and variation from galaxy to galaxy
in determining what fraction of the clusters in a starburst is visible.
For example, all the data point toward a generally dustier environment
in NGC~5248 than in NGC~1512. 
This suggests
that, on an individual galaxy basis, it is difficult to 
construct a coherent picture of stellar populations, gas content, and
dust properties, based on measurements integrated over large areas of 
a starburst (as is unavoidable when angular resolution is limited, or
a galaxy is distant). 

This situation has led in the past to challenges in explaining
self-consistently various subsets of spatially integrated measurements, 
such as the UV spectral
slope, the \Hans/H$\beta$ intensity ratio, and the IR luminosity, both in
nearby starbursts (e.g., Calzetti, Kinney,
\& Storchi-Bergmann 1994, 1996; Meurer et al. 1995, 1997), and in distant ones
(e.g., Pettini et al. 1998; Meurer, Heckman, \& Calzetti 1999).
The common explanations have been that starbursts have a relatively
grey extinction curve, or that the young stars and the gas have different
spatial distribution. 
Takagi, Arimoto, \& Vansevicius (1999) and Gordon et al. (2000) 
have addressed this problem by considering also IR and sub-mm measurements,
compared with stellar synthesis models that include the effects of 
mixed dust in various geometries. 
Charlot \& Fall (2000) have extended those studies by
accounting for absorption by various
interstellar components, and limiting the lifetimes of some of the components
 as a result of their destruction by the new stars. They find they can
reproduce the mean properties of starburst galaxies if the extinction law
is grey, or if the extinction is steep but the dust has a patchy distribution.
The data we have presented clearly support the latter picture.

We have measured the properties of the compact clusters in the rings, and
analyzed them by 
means of traditional color-color diagrams, but also using explicit 
model fitting that exploits all the available information over the  five
continuum
bands we have sampled. This improves our ability to separate
the effects of age and extinction. Given the observational errors
and the remaining age-extinction degeneracy in some cases, 
most of the SEDs are well fit by a range of
spectral synthesis models. However, some of the brightest clusters reveal
an IR excess that is not reproduced by any model. In most objects, the
excess shows up only in one band, at $1.6~\mu$m, and hence it is difficult
to constrain its properties. In one object, where the IR excess is particularly
large, and there is some excess emission in the $I$ band as well, we have shown
that the excess can be fit with a roughly 2000~K thermal spectrum. Since the
bulk of the stellar radiation is from stars with effective temperatures
of $\sim 20000$~K, if the excess is the result of thermal reradiation by dust,
the dust particles will be at no more than about 100 stellar radii, i.e.,
they are  circumstellar dust shells. This result needs to be confirmed
with additional observations, especially at longer IR wavelengths, which would
probe the shape of the excess spectrum.

  We find that the brightest clusters in both 
galaxies have ages of a few Myr, and undergo only mild extinction.
The typical extinctions are $A_V=0.5$ mag in NGC~1512 and  $A_V=1$ mag
in NGC~5248. The extinction of clusters thus appears to be an 
``on/off'' phenomenon, with the clusters either completely hidden
in their birth clouds (and revealed only via the strong line emission
from the clouds), or almost completely exposed. This shows
that the stellar and supernova winds which presumably clear out the gas
are extremely efficient in clearing out the gas on timescales $\ltorder 1$~Myr.

The SEDs are well fit when assuming
 a steep, Galactic extinction law, consistent with the conclusion, above,
regarding the form of the extinction law.
Our best-fit models to the observed cluster SEDs suggest the presence also 
of older clusters, aged up to 300 Myr. However, given the uncertainties, 
almost none of these
clusters can be ruled out from instead being 20~Myr or younger. 

The clusters have a $V$-band luminosity function of the form $N(L) dL\propto
L^{-2}dL$, similar to that previously measured in the UV for these galaxies
and to what is generally found for clusters in starburst environments.
We have pointed out an ``age bias,'' where the steep luminosity function
and the fading of clusters with time combine to make young clusters
appear overabundant. The small fraction of old clusters
indicated by the best-fit models is consistent with expectations if the
starbursts in the rings are actually continuous, rather than episodic.

Evidence that other ring starbursts are episodic, rather than continuous, has been
presented before, using similar age dating arguments (e.g., Smith et al. 1999;
P\'erez-Ram\'\i rez et al. 2000). Buta et al. (2000) have emphasized the possible detection
of {\it some} old ($> 50$ Myr) clusters in the ring in NGC~1326. From 
the theoretical perspective, Elmegreen (1994)
has shown how a long period of gas accretion onto the region of the ring can
lead to a synchronized starburst when a critical gas density is reached,
and which may last only briefly, until the accreted gas is depleted.
We caution that the age bias 
needs to be taken into account when deducing the star-forming history of
a galaxy from its cluster age distribution. In the case of the galaxies
analyzed here, because we cannot rule out the near absence of old clusters,
we cannot exclude, based on the cluster ages,
 either of the possibilities, continuous or episodic, or the possibility that
clusters are destroyed before they have a chance to age.

A statistical argument against episodic star formation can, however, be made. 
The UV imaging survey of Maoz et al. (1996) included 
an unbiased selection of 110
nearby galaxies from a complete sample. In the observed sample,
71 galaxies were spirals of Hubble type Sc or earlier, and 52 were classified
as barred (SB or SAB) by de Vaucoulers et al. (1991). In this sample,
 Maoz et al.
 found five examples of UV-bright circumnuclear 
rings. Accounting for incompleteness due to the limited field of view
raised this number to seven, or about 10\% of the Sc, and earlier, spirals.
 If circumnuclear starbursts are episodic,
the models with the lowest acceptable ages suggest that 
the UV-bright phase lasts 10-20~ Myr, 
while the interval between  successive bursts is $\gtorder 100$ Myr.
This would mean that most, or all, early-type spiral galaxies undergo 
occasional circumnuclear starbursts.  While this possibility 
is not completely outlandish, it is contrary to the idea that circumnuclear
rings require special conditions that are found only in some galaxies,
which have strong bars and Lindblad resonances that channel gas accretion
onto the rings (e.g., Elmegreen 1994).

The extensive broad-band data presented here, 
combined with our modeling, allow us to derive
approximate masses and mass functions for the clusters. The mass function
in the two galaxies is consistent with the $N(m) dm\propto
m^{-2}dm$ form that has been found in the only other starburst system, 
the Antennae galaxies, where this
has been reliably measured (Zhang \& Fall 1999). 
This power-law distribution is distinct from 
the log-normal distribution observed for old
globular clusters, which has a characteristic mass scale. 
If SSCs are the progenitors of globular clusters, some evolutionary 
process, such as evaporation or tidal disruption by galactic gravitational fields,
must reshape the mass functions (Elmegreen \& Efremov 1997;
Fall \& Zhang 2001).
 
The most massive clusters
in NGC~1512 and NGC~5248 have masses in the range $10^4$ to $10^5 M_{\odot}$.
After accounting for statistics and differences in model parameters,
this is similar to the masses in the Antennae. Combined with the 
other SSC characteristics we have evaluated, there is a remarkable
uniformity in the properties of young clusters in these two different
starburst environments.

We have estimated the sizes of the brighter, isolated SSCs, and argued
that in some cases they are only marginally resolved, or unresolved,
corresponding to Gaussian radii smaller than a few pc. 
Some previous size measurements
of SSCs in other galaxies have probably not taken full account of the
systematics which can artificially broaden the image of a point source.
It is also possible that SSCs in different environments have different sizes. 
It is easy to show (e.g., Maoz et al. 1996) that the virial radius, $r$, 
the mass, $m$, and the age, $t$,
together determine whether or not a cluster is gravitationally bound, 
where, for a bound cluster,
$$
\left({t}\over{10^5~{\rm yr}}\right)^{-1}
 \left({r}\over{\rm pc}\right)^{3/2}
 \left({m}\over{10^4~M_{\odot}}\right)^{-1/2}
< 1.
$$
The parameters we have derived for most of the brighter
clusters establish them as bound objects, unless our mass estimates
are systematically high due to, e.g., an extremely top-heavy stellar IMF.  
Visual-band size measurements, such as above,
 are also an essential ingredient in the more robust 
kinematic estimates of SSC masses based on optical spectroscopy
(Ho \& Filippenko 1996a,b; Smith \& Gallagher 2000;
Sternberg 1998).

To summarize, our study of the starburst rings in NGC~1512 and
NGC~5248 has led to the following main conclusions.
\begin{enumerate}
\item{The young star clusters and the line-emitting gas have different 
spatial distributions. The gas is mixed with dust, making it optically thick,
and effective at completely obscuring some of the young clusters, which are
revealed only by the line emission from the gas. 
}\item{ Most of the clusters that are visible are only mildly
reddened, suggesting that the processes that clear out the gas and dust
are efficient and fast.
}\item{ Some of the brightest young clusters have an IR excess that may be
thermal reradiation by circumstellar dust. 
}\item{ The cluster SEDs are consistent with a range in ages, from 1~Myr to
300~Myr. Although only a minority appears to be old, an age bias that causes
young clusters to be over-represented makes this fraction consistent with
continuous star formation in the rings over the past $\sim 300$~Myr. Due to the uncertainties in dating the clusters, we cannot rule out episodic,
$\sim 20$~Myr-long bursts of star formation, although the statistics of the 
occurrence of UV-bright nuclear rings in spiral galaxies argue against this possibility. A third possibility is that old clusters are, in fact, absent, but
due to the destruction of clusters after $\sim 20$~Myr. 
}\item{ The sizes of the clusters are of order a few pc, or less. Considering 
also the range of ages and the large masses we have established for the 
clusters, this confirms that they are gravitationally bound objects.
}\item{ The luminosity functions and the mass functions follow a power-law
distribution with index $-2$, as seen in other starburst environments. The lack
of a mass scale means that subsequent evolution of the mass
function is required, if some of the SSCs are to evolve into globular clusters.
}
\item {There is remarkable similarity in the 
 properties of the gas, dust, and clusters in different starbursts that 
have been studied in detail, whether in circumnuclear rings or in merging
galaxies.
}\item{ In NGC~5248, there is a previously unknown, inner, emission-line ring 
of radius 60~pc. In NGC~1512, we have found a peculiar compact emission-line
source with $\sim 7000$~\AA~ \Hans+[N~II] equivalent width, which
may be a young, Balmer-dominated supernova remnant.}
\end{enumerate}

\acknowledgements
We thank A. Pauldrach and R.-P. Kudritzki for providing us with
their model atmospheres in advance of publication, and the anonymous
referee for constructive comments.
This work was funded in part by grants GO-6738 and GO-7879 from the Space Telescope
Science Institute, which is operated by AURA, Inc., under NASA contract NAS
5-26555.  D.~M. and A.~S. acknowledge support by a grant from the Israel Science
Foundation. Research by A.~J.~B. is supported by a 
postdoctoral fellowship from the Harvard-Smithsonian Center for
Astrophysics. 
A.~S. acknowledges support by German-Israeli Foundation grant
I-551-186.07/97.
A.~V.~F. and L.~C.~H. acknowledge funding from NASA grant NAG5-3556.

\begin{deluxetable}{cccccccc}
\footnotesize
\tablewidth{0pt}
\tablecaption{Observation Log}
\tablehead{\colhead {Instrument} & \colhead {Filter} & \colhead{Description}&
\colhead {Inv. Sensitivity} &\multicolumn{2}{c}{NGC 1512}&\multicolumn{2}{c}{NGC 5248}
\nl
\colhead{\& Scale}&\colhead{}&\colhead{}&\colhead{}&
\colhead {Exposure} & \colhead
 {UT Date} &\colhead {Exposure} & \colhead
 {UT Date} \nl}
\startdata

WFPC2   &F336W& ``$U$''&$5.613\times10^{-17}$&2500&05/03/99&2300&30/01/99\nl
$0\asecp0455$ pix$^{-1}$        &F547M& ``$V$''&$7.691\times10^{-18}$&900 &05/03/99& 900&30/01/99\nl
        &F814W& ``$I$''&$2.508\times10^{-18}$&1100&05/03/99&1100&30/01/99\nl
        &F658N& H$\alpha$+[N~II]&$4.187\times10^{-15}$&5200&05/03/99&4800&30/01/99\nl
\hline
NICMOS-2&F160W& ``$H$''&$2.406\times10^{-19}$&2560&29/07/98&2304&24/08/98\nl
$0\asecp0764$ pix$^{-1}$       &F187W& Pa$\alpha$ cont.&$3.184\times10^{-19}$&2560&29/07/98&2304&24/08/98\nl
        &F187N& Pa$\alpha$&$7.534\times10^{-16}$&2560&29/07/98&\nodata&\nodata\nl
        &F187N& Pa$\alpha$&$9.727\times10^{-16}$&\nodata&\nodata&2304&24/08/98\nl
\enddata	
\tablecomments{Inverse sensitivity 
is in units  of erg s$^{-1}$ cm$^{-2}$ ~\AA$^{-1}$ per count s$^{-1}$ for the broad and medium band filters,
and in erg s$^{-1}$ cm$^{-2}$ per count s$^{-1}$ for the two narrow-band filters (F658N and F187N). \Pa falls on different parts of the bandpass in the two
galaxies, resulting in different sensitivities. Exposure times in seconds.
UT date format is DD/MM/YY.}
\end{deluxetable}

\begin{deluxetable}{rrrrrcccccc}
\tiny
\tablewidth{0pt}
\tablecaption{Emission-Line Complexes}
\tablehead{\colhead {No.} & \colhead {X} & \colhead {Y} & \colhead {$\Delta\alpha['']$} &
\colhead {$\Delta\delta['']$} & \colhead {Radius} & \colhead
 {$f({\rm H}\alpha)$} &
\colhead {$f({\rm Pa}\alpha)$} &
\colhead {$f_{\lambda}(1.6~ \mu{\rm m}$)} &
 \colhead {$f({\rm H}\alpha)/f({\rm Pa}\alpha)$} &
\colhead {$f({\rm Pa}\alpha)/f_{\lambda}(1.6~ \mu{\rm m}$)} \nl
\colhead {(1)}&\colhead {(2)}&\colhead {(3)}&\colhead {(4)}&
\colhead {(5)}&\colhead {(6)}&\colhead {(7)}&\colhead {(8)}&\colhead {(9)}&
\colhead {(10)}&\colhead {(11)}\nl}
\startdata
&&&&&&&{\bf NGC 1512}&&&\nl
\hline

 1&446.&549.&  5.53& $-$4.77&$0\asecp$68& 179.6&44.4& 31.0& 4.0&143\nl
 2&405.&534.&  3.66& $-$5.44&$0\asecp$68& 255.2&50.5& 59.2& 5.1& 85\nl
 3&352.&515.&  1.25& $-$6.32&$0\asecp$68& 289.9&58.8& 39.5& 4.9&149\nl
 4&299.&450.& $-$2.49& $-$5.58&$0\asecp$45&  91.8&24.1& 23.3& 3.8&103\nl
 5&275.&434.& $-$3.80& $-$5.72&$0\asecp$45&  88.7&24.1& 21.9& 3.7&110\nl
 6&263.&352.& $-$6.60& $-$3.20&$0\asecp$23&  16.9& 9.8&  3.6& 1.7&271\nl
 7&324.&243.& $-$7.64&  2.39&$0\asecp$23&  18.5& 4.5&  1.7& 4.1&268\nl
 8&361.&234.& $-$6.61&  3.78&$0\asecp$45&  70.3&24.1&  7.7& 2.9&313\nl
 9&574.&394.&  5.50&  4.38&$0\asecp$45&  69.2& 8.3& 17.6& 8.4& 47\nl
10&577.&428.&  6.59&  3.27&$0\asecp$45&  46.5& 8.3& 13.2& 5.6& 63\nl
11&540.&536.&  8.44& $-$1.58&$0\asecp$45&  51.0&12.1&  9.4& 4.2&128\nl
\hline
&&&&&&&{\bf NGC 5248}&&&\nl
\hline
 1&310.&449.&  3.25&  4.61&$0\asecp$32&  30.6&10.7&  9.6& 2.9&111\nl
 2&333.&421.&  2.97&  2.99&$0\asecp$18&  16.9& 8.8&  5.9& 1.9&149\nl
 3&277.&421.&  5.19&  4.24&$0\asecp$18&   1.2& 4.9&  1.3& 0.2&361\nl
 4&269.&411.&  5.73&  4.03&$0\asecp$18&   1.9& 6.8&  1.3& 0.3&515\nl
 5&272.&392.&  6.04&  3.21&$0\asecp$18&  17.0& 6.8&  1.9& 2.5&349\nl
 6&342.&375.&  3.65&  0.96&$0\asecp$27&  35.2& 8.8& 18.6& 4.0& 47\nl
 7&329.&359.&  4.52&  0.62&$0\asecp$32&  35.8& 9.7& 23.9& 3.7& 41\nl
 8&342.&337.&  4.50&  $-$.54&$0\asecp$23&  27.1& 3.9& 14.3& 7.0& 27\nl
 9&228.&285.& 10.18&  $-$.05&$0\asecp$18&   7.8& 6.8&  0.7& 1.1&943\nl
10&310.&289.&  6.84& $-$1.72&$0\asecp$18&  12.1& 4.9&  3.3& 2.5&146\nl
11&346.&285.&  5.50& $-$2.69&$0\asecp$27&  17.5&11.7& 13.1& 1.5& 89\nl
12&386.&277.&  4.10& $-$3.90&$0\asecp$45& 104.7&25.3& 29.1& 4.1& 87\nl
13&471.&266.&  0.98& $-$6.24&$0\asecp$32&  21.8&13.6&  8.0& 1.6&170\nl
14&509.&346.& $-$2.32& $-$3.93&$0\asecp$91& 358.0&72.0&118.9& 5.0& 61\nl
15&543.&406.& $-$5.01& $-$2.31&$0\asecp$36&  33.8&20.4& 16.8& 1.7&121\nl
16&580.&416.& $-$6.70& $-$2.75&$0\asecp$36&   8.8&21.4&  3.8& 0.4&566\nl
17&531.&481.& $-$6.21&  0.93&$0\asecp$36&  45.2&11.7& 20.3& 3.9& 58\nl
18&464.&489.& $-$3.74&  2.74&$0\asecp$23&  21.8& 6.8& 11.7& 3.2& 58\nl
19&444.&504.& $-$3.28&  3.79&$0\asecp$23&  18.7& 4.9& 10.9& 3.8& 45\nl
20&387.&564.& $-$2.37&  7.44&$0\asecp$18&  13.6& 3.9&  2.1& 3.5&186\nl
21&528.&561.& $-$7.89&  4.16&$0\asecp$18&  33.6& 6.8&  8.5& 4.9& 80\nl
\enddata	
\tablecomments{(2)(3) X and Y 
PC pixel coordinates; (4)(5) Offset in arcseconds from the galaxy nucleus
(in NGC 1512: pixel coordinates 419.6, 390.7; in NGC~5248:
pixel coordinates 422.16, 396.02), in the R.A. and Decl. directions; (6) Angular radius of 
circular aperture used for flux measurements; at a distance of 10 Mpc,
$0.1''$ corresponds to 5 pc; (7)(8)
Line fluxes in units of $10^{-16}$ erg s$^{-1}$ cm$^{-2}$. At a 
distance of 10 Mpc this unit
corresponds to a luminosity of $1.2\times10^{36}$ erg s$^{-1}$.
\Ha denotes \Hans+[N~II] flux in the F658N filter, as explained in text;
(9) $f_{\lambda}(1.6~\mu{\rm m}$)
in units  of $10^{-18}$ erg s$^{-1}$ cm$^{-2}$ \AA$^{-1}$;
(11) $f({\rm Pa}\alpha)/f_{\lambda}(1.6~ \mu{\rm m}$) in \AA.}
\end{deluxetable}

\begin{deluxetable}{rrrrrcccccccccccc}
\tiny
\tablewidth{0pt}
\tablecaption{Compact Sources Brightest in $V$ -- NGC~1512
}
\tablehead{
\colhead {No.} & 
\colhead {X} & 
\colhead {Y} &
\colhead {$\Delta\alpha$} &
\colhead {$\Delta\delta$} & 
\colhead {$V$}&
\colhead {$\Delta_{1-3}$}&
\colhead {$f_{\lambda}$} &\colhead{$\sigma$}&
\colhead {$f_{\lambda}$} &\colhead{$\sigma$}&
\colhead {$f_{\lambda}$} &\colhead{$\sigma$}&
\colhead {$f_{\lambda}$} &\colhead{$\sigma$}&
\colhead {$f_{\lambda}$} &\colhead{$\sigma$} \nl
\colhead {} & 
\colhead {PC} & 
\colhead {PC} &
\colhead {[$''$]} &
\colhead {[$''$]} & 
\colhead {mag}&
\colhead {mag}&
\colhead {$2300~$\AA} &\colhead{}&
\colhead {$3350~$\AA} &\colhead{}&
\colhead {$5490~$\AA} &\colhead{}&
\colhead {$8040~$\AA} &\colhead{}&
\colhead {$1.6 \mu{\rm m}$}&\colhead{} \nl
\colhead {(1)}&\colhead {(2)}&\colhead {(3)}&\colhead {(4)}&
\colhead {(5)}&\colhead {(6)}&\colhead {(7)}&\colhead {(8)}&\colhead {(9)}&
\colhead {(10)}&\colhead {(11)}&\colhead {(12)}&\colhead {(13)}&\colhead {(14)}&
\colhead {(15)}&\colhead {(16)}&\colhead {(17)}\nl}
\startdata
  1&  407.42&  534.64&    3.76&   -5.39& 19.10&  0.98& 22000.0&  4400.0&  2957.5& 149.8&   755.9&  38.9&  237.0&  12.4&    39.9&    4.0\nl
  2&  350.21&  512.03&    1.10&   -6.26& 20.22&  1.14&  3450.0&   690.0&  1212.1&  61.6&   269.3&  14.0&   85.2&   4.6&    13.8&    1.2\nl
  3&  444.00&  549.25&    5.47&   -4.84& 20.31&  1.05&  1650.0&   330.0&   868.9&  44.1&   248.2&  13.0&   86.4&   4.8&    15.7&    1.8\nl
  4&  277.43&  434.07&   -3.71&   -5.65& 20.68&  1.15&  1360.0&   272.0&   555.7&  30.6&   176.4&   9.7&   70.8&   4.1&    14.4&    2.5\nl
  5&  398.78&  528.47&    3.28&   -5.43& 20.74&  1.33&   670.0&   134.0&   458.1&  28.2&   167.2&  10.2&   79.7&   5.0&    59.0&    4.3\nl
  6&  250.51&  414.17&   -5.23&   -5.74& 21.07&  0.86&   210.0&    42.0&   177.6&   9.6&   122.9&   6.6&   62.5&   3.3&    16.5&    1.6\nl
  7&  405.49&  531.97&    3.62&   -5.35& 21.19&  3.09&   $<30$&    \nod&   488.4&  29.8&   110.3&   8.6&   50.9&   4.1&    22.6&    2.9\nl
  8&  444.72&  545.52&    5.38&   -4.69& 21.26&  1.57&  1000.0&   200.0&   432.6&  23.1&   103.4&   6.4&   35.6&   3.0&    14.4&    1.6\nl
  9&  548.57&  317.11&    2.37&    6.33& 21.38&  1.06&   270.0&    54.0&   194.2&  10.8&    92.0&   5.4&   39.4&   2.1&     8.9&    0.8\nl
 10&  573.71&  392.75&    5.45&    4.41& 21.51&  1.52&   610.0&   122.0&   266.2&  14.2&    82.3&   4.8&   30.3&   2.2&     8.6&    0.9\nl
 11&  541.50&  526.25&    8.21&   -1.20& 21.64&  1.27&   300.0&    60.0&   191.7&  10.7&    72.9&   4.4&   28.5&   1.7&     7.9&    0.9\nl
 12&  257.85&  419.39&   -4.82&   -5.71& 21.78&  0.81&   200.0&    40.0&   186.7&  11.1&    64.1&   4.6&   24.2&   2.2&     4.1&    1.4\nl
 13&  258.43&  323.01&   -7.61&   -2.32& 21.82&  1.42&   $<30$&    \nod&    64.4&   4.8&    61.5&   3.8&   34.1&   2.1&    11.7&    1.1\nl
 14&  306.08&  446.24&   -2.36&   -5.25& 21.84&  1.50&   200.0&    40.0&   132.4&   8.9&    60.7&   3.7&   27.1&   1.6&     6.5&    1.3\nl
 15&  307.29&  447.80&   -2.27&   -5.26& 21.89&  1.57&   370.0&    74.0&   146.3&   8.7&    57.8&   3.4&   23.5&   1.4&     3.4&    1.3\nl
 16&  269.34&  423.21&   -4.31&   -5.51& 21.95&  1.05&   190.0&    38.0&   119.2&   9.3&    54.6&   4.6&   27.0&   2.5&    21.9&    2.0\nl
 17&  353.49&  512.40&    1.23&   -6.18& 21.98&  3.14&   $<30$&    \nod&   214.9&  13.0&    53.2&   4.2&   25.3&   2.3&     9.6&    1.0\nl
 18&  272.16&  427.69&   -4.08&   -5.58& 22.09&  1.50&   240.0&    48.0&   151.2&  10.8&    48.0&   4.2&   15.1&   2.1&    14.7&    2.0\nl
 19&  300.79&  446.65&   -2.53&   -5.41& 22.12&  1.04&   320.0&    64.0&   185.1&  10.8&    46.6&   3.8&   14.0&   1.3&     2.5&    1.2\nl
 20&  256.38&  321.62&   -7.72&   -2.33& 22.14&  1.60&   $<30$&    \nod&    72.5&   5.1&    45.9&   3.1&   27.8&   1.6&     9.2&    1.0\nl
 21&  296.04&  454.57&   -2.46&   -5.83& 22.14&  1.05&   270.0&    54.0&   135.5&   8.3&    45.7&   3.5&   16.4&   2.4&    $<1$&   \nod\nl
 22&  439.91&  196.02&   -4.95&    7.40& 22.16&  0.85&   $<30$&    \nod&   305.2&  15.8&    45.3&   2.6&   17.1&   1.0&    $<1$&   \nod\nl
 23&  244.98&  404.33&   -5.71&   -5.56& 22.16&  1.01&   $<30$&    \nod&    60.5&   4.1&    45.1&   2.9&   18.8&   1.2&     7.7&    1.4\nl
 24&  439.87&  534.85&    4.90&   -4.45& 22.18&  1.29&   $<30$&    \nod&    60.0&   6.7&    44.2&   3.6&   34.6&   2.5&    30.7&    1.8\nl
 25&  280.84&  451.00&   -3.10&   -6.15& 22.20&  1.12&    90.0&    18.0&    94.2&   6.8&    43.5&   3.5&   15.0&   1.8&     4.8&    1.5\nl
 26&  593.89&  475.05&    8.55&    2.12& 22.20&  1.46&   140.0&    28.0&    74.1&   5.3&    43.5&   3.0&   23.9&   1.6&    11.9&    1.2\nl
 27&  443.00&  534.75&    5.01&   -4.36& 22.21&  1.59&   $<30$&    \nod&    35.1&   5.8&    43.1&   3.7&   33.3&   2.3&    25.9&    1.9\nl
 28&  402.32&  540.55&    3.76&   -5.75& 22.25&  1.64&   $<30$&    \nod&   125.0&   7.9&    41.5&   3.3&   18.0&   1.8&    13.7&    1.2\nl
 29&  603.75&  721.68&   16.07&   -6.22& 22.31&  1.15&    \nod&    \nod&    19.7&   2.3&    39.2&   2.3&   28.8&   1.5&    \nod&   \nod\nl
 30&  597.08&  511.50&    9.72&    0.94& 22.31&  0.95&   $<30$&    \nod&     5.7&   1.3&    39.1&   2.4&   40.0&   2.1&    21.2&    1.2\nl
\enddata	
\tablecomments{
The complete version of this table is in the electronic edition of
the Journal. The printed edition contains only a sample;
(2)(3) X and Y 
pixel PC coordinates in the F547M frame; 
(4)(5) Offset in arcseconds from the galaxy nucleus
(at pixel coordinates 419.6, 390.7) 
in the R.A. and Decl. directions; 
(6) $V$ magnitude;
(7) Magnitude difference 
between F547M photometry in 3-pixel- and 1-pixel-radius apertures;
(8)-(17) $f_{\lambda}$ and $1\sigma$ errors
in units  of $10^{-19}$ erg s$^{-1}$ cm$^{-2}$ \AA$^{-1}$.
Missing entries are for sources outside the field of view of the
FOC or NICMOS. For null detections, the
flux limits actually depend on the local background and crowding, and the
limits quoted are just those typical for the band.}
\end{deluxetable}

\begin{deluxetable}{rrrrrcccccccccccc}
\tiny
\tablewidth{0pt}
\tablecaption{Compact Sources Brightest in $V$ -- NGC~5248}
\tablehead{
\colhead {No.} & 
\colhead {X} & 
\colhead {Y} &
\colhead {$\Delta\alpha$} &
\colhead {$\Delta\delta$} &
\colhead {$V$}& 
\colhead {$\Delta_{1-3}$}&
\colhead {$f_{\lambda}$} &\colhead{$\sigma$}&
\colhead {$f_{\lambda}$} &\colhead{$\sigma$}&
\colhead {$f_{\lambda}$} &\colhead{$\sigma$}&
\colhead {$f_{\lambda}$} &\colhead{$\sigma$}&
\colhead {$f_{\lambda}$} &\colhead{$\sigma$} \nl
\colhead {} & 
\colhead {PC} & 
\colhead {PC} &
\colhead {[$''$]} &
\colhead {[$''$]} & 
\colhead {mag}&
\colhead {mag}&
\colhead {$2300$~\AA} &\colhead{}&
\colhead {$3350$~\AA} &\colhead{}&
\colhead {$5490$~\AA} &\colhead{}&
\colhead {$8040$~\AA} &\colhead{}&
\colhead {$1.6 \mu{\rm m}$}&\colhead{} \nl
\colhead {(1)}&\colhead {(2)}&\colhead {(3)}&\colhead {(4)}&
\colhead {(5)}&\colhead {(6)}&\colhead {(7)}&\colhead {(8)}&\colhead {(9)}&
\colhead {(10)}&\colhead {(11)}&\colhead {(12)}&\colhead {(13)}&\colhead {(14)}&
\colhead {(15)}&\colhead {(16)}&\colhead {(17)}\nl}
\startdata
  1&  527.45&  560.56&   -7.86&    4.16& 18.26&  0.91&  2810.0&   562.0&  3510.9& 176.4&  1641.9&  82.5&  649.2&  32.7&    84.7&    4.7\nl
  2&  302.41&  311.83&    6.63&   -0.65& 19.64&  0.84&   300.0&    60.0&   621.6&  31.8&   460.1&  23.8&  209.9&  11.0&    28.5&    3.8\nl
  3&  330.29&  312.58&    5.51&   -1.25& 19.87&  1.10&  1110.0&   222.0&   806.3&  42.3&   370.5&  19.8&  168.2&   9.4&   108.4&    7.7\nl
  4&  280.13&  324.35&    7.23&    0.34& 20.15&  0.97&   530.0&   106.0&   522.4&  27.0&   287.8&  14.8&  129.9&   6.7&    44.9&    3.4\nl
  5&  411.89&  262.85&    3.39&   -5.04& 20.29&  1.38&   660.0&   132.0&   496.2&  28.4&   252.9&  15.7&  132.0&   8.4&    89.8&    7.6\nl
  6&  498.44&  339.01&   -1.74&   -3.97& 20.41&  1.18&   520.0&   104.0&   442.7&  23.4&   225.0&  12.0&  102.9&   5.8&    30.3&    2.3\nl
  7&  513.45&  348.56&   -2.55&   -3.92& 20.48&  1.25&  1220.0&   244.0&   733.9&  38.8&   211.1&  12.1&   72.0&   4.2&    17.1&    2.3\nl
  8&  554.06&  478.15&   -7.06&    0.30& 20.66&  1.01&   770.0&   154.0&   511.7&  26.9&   179.2&   9.9&   61.1&   3.6&     8.4&    1.8\nl
  9&  454.36&  520.05&   -4.05&    4.19& 20.71&  1.02&   430.0&    86.0&   447.5&  23.4&   171.5&   9.5&   63.2&   3.9&    15.3&    2.6\nl
 10&  502.67&  343.70&   -2.02&   -3.88& 20.77&  1.01&   230.0&    46.0&   337.3&  19.2&   162.2&   9.6&   69.4&   4.9&     9.8&    1.6\nl
 11&  346.77&  268.92&    5.83&   -3.34& 20.82&  1.16&   260.0&    52.0&   294.2&  15.6&   154.8&   8.8&   73.8&   4.5&    34.3&    3.8\nl
 12&  338.38&  187.17&    8.00&   -6.39& 20.88&  1.01&   410.0&    82.0&   327.8&  17.3&   146.2&   7.8&   58.3&   3.2&    18.4&    1.4\nl
 13&  344.61&  375.92&    3.52&    0.94& 20.93&  1.23&   350.0&    70.0&   333.2&  17.7&   140.3&   8.5&   52.1&   4.0&     5.7&    1.7\nl
 14&  515.11&  349.60&   -2.64&   -3.92& 20.96&  1.92&   $<30$&    \nod&   422.3&  24.2&   136.3&   9.1&   55.1&   4.8&    18.5&    2.7\nl
 15&  504.04&  342.91&   -2.05&   -3.94& 21.10&  2.26&   230.0&    46.0&   242.4&  14.8&   119.2&   7.8&   53.6&   3.9&    12.9&    1.7\nl
 16&  496.68&  336.63&   -1.62&   -4.02& 21.17&  1.96&   $<30$&    \nod&   218.2&  12.2&   112.3&   6.4&   62.1&   3.8&    16.2&    2.1\nl
 17&  535.25&  481.75&   -6.40&    0.86& 21.18&  1.60&   240.0&    48.0&   233.5&  13.5&   110.7&   7.2&   41.2&   3.6&    13.3&    1.9\nl
 18&  341.92&  538.23&   -0.01&    7.43& 21.26&  1.13&   240.0&    48.0&   191.3&  10.6&   103.0&   6.0&   56.4&   3.3&    35.5&    2.5\nl
 19&  531.00&  570.30&   -8.22&    4.46& 21.28&  1.00&   320.0&    64.0&   231.4&  15.1&   101.7&   8.0&   42.6&   3.9&    \nod&   \nod\nl
 20&  332.91&  538.85&    0.33&    7.66& 21.29&  1.24&   190.0&    38.0&   161.3&   9.2&   100.6&   5.5&   59.2&   3.3&    27.0&    2.2\nl
 21&  525.62&  347.69&   -3.01&   -4.23& 21.35&  1.06&   140.0&    28.0&   191.8&  12.8&    94.8&   7.8&   39.3&   4.0&    26.5&    3.7\nl
 22&  341.11&  318.68&    4.94&   -1.25& 21.39&  1.50&   $<30$&    \nod&   150.5&  10.9&    91.4&   7.6&   46.9&   4.9&    23.2&    4.9\nl
 23&  280.47&  340.22&    6.86&    0.97& 21.42&  1.28&   220.0&    44.0&   158.7&   8.9&    89.1&   5.2&   49.3&   3.0&    16.5&    3.1\nl
 24&  342.77&  364.74&    3.84&    0.54& 21.45&  1.42&   $<30$&    \nod&   136.4&  14.4&    86.5&  12.0&   35.1&   6.0&    29.8&    2.7\nl
 25&  341.16&  374.71&    3.68&    0.97& 21.48&  1.55&   160.0&    32.0&   227.8&  13.2&    84.3&   6.3&   36.5&   3.4&     9.5&    2.0\nl
 26&  342.99&  374.29&    3.62&    0.91& 21.50&  2.42&   350.0&    70.0&   206.2&  12.4&    82.9&   6.7&   40.0&   3.9&    11.8&    1.8\nl
 27&  443.86&  379.88&   -0.50&   -1.13& 21.55&  2.03&   $<30$&    \nod&    67.9&   5.8&    79.0&   8.1&   52.4&   6.9&    $<1$&   \nod\nl
 28&  431.58&  376.18&    0.07&   -1.00& 21.56&  1.72&   $<30$&    \nod&    63.1&   4.9&    78.2&   6.6&   58.2&   6.4&    $<1$&   \nod\nl
 29&  495.80&  498.68&   -5.22&    2.42& 21.59&  1.34&   230.0&    46.0&   177.0&  10.1&    76.5&   5.2&   31.8&   2.7&     9.9&    1.7\nl
 30&  407.68&  264.10&    3.53&   -4.90& 21.59&  1.49&   $<30$&    \nod&    90.7&  13.9&    76.4&   9.3&   55.7&   4.8&    60.5&    5.9\nl
\enddata               
\tablecomments{
The complete version of this table is in the electronic edition of
the Journal. The printed edition contains only a sample;
(4)(5) Offset in arcseconds from the galaxy nucleus
(at                    
pixel coordinates 422.16, 396.02) in the R.A. and Decl. directions; 
other notes as in Table 3. 
}
\end{deluxetable}      

\begin{deluxetable}{rrcccccccccccc}
\tiny
\tablewidth{0pt}
\tablecaption{Cluster Ages, Extinctions, and Masses -- NGC~1512}
\tablehead{
\colhead{}&
\colhead{}&
\multicolumn{4}{c}{Best Fittting Model}&
\multicolumn{4}{c}{Youngest Acceptable Model}&
\multicolumn{4}{c}{Oldest Acceptable Model}\nl
\colhead {No.} & 
\colhead {Bands}& 
\colhead {Age} & 
\colhead {$A_V$} &
\colhead {Mass} &
\colhead {$\chi^2$/dof} &
\colhead {Age} & 
\colhead {$A_V$} &
\colhead {Mass} &
\colhead {$\chi^2$/dof} &
\colhead {Age} & 
\colhead {$A_V$} &
\colhead {Mass} &
\colhead {$\chi^2$/dof} \nl
\colhead {} &
\colhead {} &
\colhead {Myr} & 
\colhead {mag} &
\colhead {$10^4M_{\odot}$} &
\colhead {} &
\colhead {Myr} & 
\colhead {mag} &
\colhead {$10^4M_{\odot}$} &
\colhead {} &
\colhead {Myr} & 
\colhead {mag} &
\colhead {$10^4M_{\odot}$} &
\colhead {}\nl
}
\startdata
   1& 5&   1&  0.0& 9.08&99.59&\nod& \nod& \nod& \nod&\nod& \nod& \nod& \nod\nl
   2& 5&   3&  0.2& 0.72& 8.55&\nod& \nod& \nod& \nod&\nod& \nod& \nod& \nod\nl
   3& 5&   1&  0.6& 3.72& 4.45&\nod& \nod& \nod& \nod&\nod& \nod& \nod& \nod\nl
   4& 5&   5&  0.1& 0.31& 2.52&\nod& \nod& \nod& \nod&\nod& \nod& \nod& \nod\nl
   5& 5&  10&  0.0& 0.51&15.42&\nod& \nod& \nod& \nod&\nod& \nod& \nod& \nod\nl
   6& 5&  20&  0.6& 1.35& 1.61&  20&  0.6& 1.35& 1.61&  20&  0.6& 1.35& 1.61\nl
   7& 4&   3&  0.6& 0.49&17.59&\nod& \nod& \nod& \nod&\nod& \nod& \nod& \nod\nl
   8& 5&   1&  0.4& 1.35&13.17&\nod& \nod& \nod& \nod&\nod& \nod& \nod& \nod\nl
   9& 5&   5&  0.6& 0.24& 2.14&\nod& \nod& \nod& \nod&\nod& \nod& \nod& \nod\nl
  10& 5&   5&  0.1& 0.14& 7.27&\nod& \nod& \nod& \nod&\nod& \nod& \nod& \nod\nl
  11& 5&  20&  0.1& 0.55& 3.22&\nod& \nod& \nod& \nod&\nod& \nod& \nod& \nod\nl
  12& 5&   1&  1.0& 1.36& 0.35&   1&  0.9& 1.12& 1.26&   3&  0.9& 0.29& 0.45\nl
  13& 4&  20&  1.0& 0.94& 1.29&   7&  0.5& 0.13& 1.80& 100&  0.0& 1.18& 1.36\nl
  14& 5&  20&  0.2& 0.48& 1.25&   5&  0.5& 0.14& 1.63&  20&  0.2& 0.48& 1.25\nl
  15& 5&   5&  0.2& 0.11& 0.53&   5&  0.2& 0.11& 0.53&   5&  0.2& 0.11& 0.53\nl
  16& 5&  12&  0.1& 0.22& 6.09&\nod& \nod& \nod& \nod&\nod& \nod& \nod& \nod\nl
  17& 4&   3&  0.8& 0.28&20.73&\nod& \nod& \nod& \nod&\nod& \nod& \nod& \nod\nl
  18& 5&  20&  0.0& 0.35& 9.92&\nod& \nod& \nod& \nod&\nod& \nod& \nod& \nod\nl
  19& 5&   3&  0.5& 0.15& 0.75&   1&  0.5& 0.60& 0.80&   3&  0.4& 0.12& 1.24\nl
  20& 4&   7&  0.3& 0.10& 0.46&   7&  0.2& 0.09& 1.09&  70&  0.0& 0.68& 1.34\nl
  21& 4&   1&  0.6& 0.61& 1.29&   1&  0.6& 0.61& 1.29&   5&  0.2& 0.08& 1.61\nl
  23& 4&  20&  0.6& 0.47& 0.97&  20&  0.5& 0.42& 1.72&  40&  0.0& 0.43& 1.95\nl
  24& 4&  12&  0.6& 0.33& 1.21&  12&  0.6& 0.33& 1.21&  12&  0.6& 0.33& 1.21\nl
  25& 5&   1&  1.2& 1.00& 1.17&   1&  1.1& 0.84& 1.77&   3&  1.1& 0.21& 1.41\nl
  26& 5&  15&  0.1& 0.20& 1.83&  15&  0.1& 0.20& 1.83&  15&  0.1& 0.20& 1.83\nl
  27& 4&  10&  0.9& 0.28& 1.73&  10&  0.9& 0.28& 1.73&  10&  0.9& 0.28& 1.73\nl
  28& 4&  20&  0.0& 0.35&21.46&\nod& \nod& \nod& \nod&\nod& \nod& \nod& \nod\nl
  30& 4& 300&  2.2&11.03& 2.04&\nod& \nod& \nod& \nod&\nod& \nod& \nod& \nod\nl
\enddata               
\tablecomments{
The complete version of this table is in the electronic edition of
the Journal. The printed edition contains only a sample;
Objects with detections in fewer than 4 bands not fitted.
Formally ``acceptable'' models have $\chi^2$ per degree of freedom $<2$.
}
\end{deluxetable}

\begin{deluxetable}{rrcccccccccccc}
\tiny
\tablewidth{0pt}
\tablecaption{Cluster Ages, Extinctions, and Masses -- NGC~5248}
\tablehead{
\colhead{}&
\colhead{}&
\multicolumn{4}{c}{Best Fittting Model}&
\multicolumn{4}{c}{Youngest Acceptable Model}&
\multicolumn{4}{c}{Oldest Acceptable Model}\nl
\colhead {No.} & 
\colhead {Bands}& 
\colhead {Age} & 
\colhead {$A_V$} &
\colhead {Mass} &
\colhead {$\chi^2$/dof} &
\colhead {Age} & 
\colhead {$A_V$} &
\colhead {Mass} &
\colhead {$\chi^2$/dof} &
\colhead {Age} & 
\colhead {$A_V$} &
\colhead {Mass} &
\colhead {$\chi^2$/dof} \nl
\colhead {} &
\colhead {} &
\colhead {Myr} & 
\colhead {mag} &
\colhead {$10^4M_{\odot}$} &
\colhead {} &
\colhead {Myr} & 
\colhead {mag} &
\colhead {$10^4M_{\odot}$} &
\colhead {} &
\colhead {Myr} & 
\colhead {mag} &
\colhead {$10^4M_{\odot}$} &
\colhead {}\nl
}
\startdata
   1& 5&   1&  1.3&41.85& 0.27&   1&  1.3&41.85& 0.27&   3&  1.2& 8.88& 1.53\nl
   2& 5&   1&  1.9&19.58& 2.52&\nod& \nod& \nod& \nod&\nod& \nod& \nod& \nod\nl
   3& 5&  40&  0.1& 4.89&13.01&\nod& \nod& \nod& \nod&\nod& \nod& \nod& \nod\nl
   4& 5&  20&  0.5& 2.93& 1.58&  20&  0.5& 2.93& 1.58&  20&  0.5& 2.93& 1.58\nl
   5& 5&  40&  0.2& 3.72& 8.59&\nod& \nod& \nod& \nod&\nod& \nod& \nod& \nod\nl
   6& 5&  20&  0.4& 2.14& 1.41&  20&  0.4& 2.14& 1.41&  20&  0.4& 2.14& 1.41\nl
   7& 5&   3&  0.6& 0.75& 4.54&\nod& \nod& \nod& \nod&\nod& \nod& \nod& \nod\nl
   8& 5&   1&  0.8& 2.96& 0.44&   1&  0.8& 2.96& 0.44&   3&  0.7& 0.62& 1.48\nl
   9& 5&   1&  1.1& 3.83& 2.05&\nod& \nod& \nod& \nod&\nod& \nod& \nod& \nod\nl
  10& 5&   3&  1.4& 1.15& 0.03&   1&  1.4& 4.59& 0.19&   3&  1.3& 0.98& 0.98\nl
  11& 5&  20&  0.6& 1.87& 3.59&\nod& \nod& \nod& \nod&\nod& \nod& \nod& \nod\nl
  12& 5&  20&  0.3& 1.30& 3.44&\nod& \nod& \nod& \nod&\nod& \nod& \nod& \nod\nl
  13& 5&   1&  1.1& 2.99& 0.23&   1&  1.0& 2.49& 1.84&   3&  1.0& 0.63& 0.81\nl
  14& 4&   5&  0.1& 0.25& 7.16&\nod& \nod& \nod& \nod&\nod& \nod& \nod& \nod\nl
  15& 5&   5&  0.8& 0.37& 1.92&   5&  0.8& 0.37& 1.92&   5&  0.8& 0.37& 1.92\nl
  16& 4&   7&  0.0& 0.18& 0.72&   7&  0.0& 0.18& 0.72&  30&  0.0& 1.03& 1.94\nl
  17& 5&  20&  0.4& 1.03& 4.11&\nod& \nod& \nod& \nod&\nod& \nod& \nod& \nod\nl
  18& 5&  40&  0.3& 1.70& 8.86&\nod& \nod& \nod& \nod&\nod& \nod& \nod& \nod\nl
  19& 4&   5&  0.5& 0.23& 1.11&   5&  0.5& 0.23& 1.11&  20&  0.2& 0.79& 1.99\nl
  20& 5&  30&  0.4& 1.32& 1.76&  30&  0.4& 1.32& 1.76&  40&  0.4& 1.72& 1.90\nl
  21& 5&  20&  0.6& 1.13& 5.91&\nod& \nod& \nod& \nod&\nod& \nod& \nod& \nod\nl
  22& 4&  40&  0.1& 1.13& 0.64&   7&  0.0& 0.14& 1.76&  70&  0.0& 1.37& 1.58\nl
  23& 5&  30&  0.2& 0.93& 1.31&  20&  0.5& 1.03& 1.70&  40&  0.2& 1.22& 1.50\nl
  24& 4&  12&  0.0& 0.30& 2.83&\nod& \nod& \nod& \nod&\nod& \nod& \nod& \nod\nl
  25& 5&   3&  1.2& 0.53& 2.02&\nod& \nod& \nod& \nod&\nod& \nod& \nod& \nod\nl
  26& 5&  20&  0.1& 0.64& 1.85&  20&  0.1& 0.64& 1.85&  20&  0.1& 0.64& 1.85\nl
  29& 5&  20&  0.3& 0.71& 2.55&\nod& \nod& \nod& \nod&\nod& \nod& \nod& \nod\nl
  30& 4&  12&  0.7& 0.60& 2.53&\nod& \nod& \nod& \nod&\nod& \nod& \nod& \nod\nl
\enddata               
\tablecomments{
The complete version of this table is in the electronic edition of
the Journal. The printed edition contains only a sample;
See Table 5.}
\end{deluxetable}      
                       
\clearpage                       
\begin{figure}         
\epsscale{.8}        
\caption{Multiwavelength view of NGC~1512.
Each panel shows a section of the image that is $18.2''$ on a
side. Arrowed lines in the F547M panel show
the orientation. The same scale and orientation apply to
the subsequent images of this galaxy, in Figs. 3-9.       
}                      
\end{figure}           
\clearpage                       
\begin{figure}         
\epsscale{.8}        
\caption{Same as Fig.1, for NGC~5248.}
\end{figure}           
\clearpage

\begin{figure}        
\epsscale{1.1}
\caption{``True color'' renditions of NGC~1512 
and NGC~5248, with blue, green, and red representing
F336W, F547M, and F814W, respectively.}                       
\end{figure}           

\clearpage
\begin{figure}        
\epsscale{1.1}
\end{figure}           

\clearpage
\begin{figure}         
\caption{Additional color composites of NGC~1512 (left) and NGC~5248 (right).
  From top to bottom:
a UV-through-IR composite, with blue, green, and red representing F220W,
F547M, and F160W, respectively;
an emission-line image, similar to the true-color
versions in Fig. 3,
but with red representing the F658N image (\Hans+[N~II]+continuum);
an IR rendition, with blue, green, and red representing
F814W, F160W, and F187W, respectively.}
\end{figure}           
\clearpage
                       
\begin{figure}         
\epsscale{1}         
\caption{Continuum-subtracted images of NGC~1512 in
\Hans+[N~II] (left) and \Pa (right). The main emission-line complexes are labeled,
and marked with the apertures used for the measurements in Table 2.
The peculiar compact  \Hans+[N~II] source discussed in the text is marked ``A''.
}                      
\end{figure}           
\clearpage                       
\begin{figure}         
\epsscale{1}         
\caption{Continuum-subtracted images of NGC~5248 in
\Hans+[N~II] (left) and \Pa (right). The main emission-line complexes are labeled,
and marked with the apertures used for the measurements in Table 2.}
\end{figure}           
\clearpage                       
\begin{figure}         
\epsscale{1}           
\caption{Left: \Hans+[N~II] emission (contours) overlaid on the F336W image
of NGC~1512 (greyscale). Right: Detail focusing on the upper left quadrant.
Note the imperfect correspondence
between the brightest \Hans+[N~II] and continuum knots.
Lowest-level contours are at 
a level of             
 $8.7\times 10^{-17}$ erg s$^{-1}$ cm$^{-2}$ arcsec$^{-2}$,
and the interior contours are at 
$2.2\times 10^{-16}$ erg s$^{-1}$ cm$^{-2}$ arcsec$^{-2}$
and 
$5.4\times 10^{-16}$ erg s$^{-1}$ cm$^{-2}$ arcsec$^{-2}$.
}                      
\end{figure}           
\clearpage                       
\begin{figure}         
\epsscale{1}           
\caption{Left: \Hans+[N~II] emission (contours) overlaid on the F336W image
of NGC~5248 (greyscale). Right: Detail. Contour levels are as in Fig. 7.} 
\end{figure}           
\clearpage                       
\begin{figure}         
\epsscale{1}           
\caption{Left: (\Hans+[N~II])/\Pa emission line ratio map for NGC~1512.
Right: Detail focusing on the upper left quadrant.
Overlaid contours are the \Pa emission, delineating the 
regions where there is significant detected flux in this line 
above a level of       
 $6\times 10^{-16}$ erg s$^{-1}$ cm$^{-2}$ arcsec$^{-2}$.
The second contour corresponds to 
$14\times 10^{-16}$ erg s$^{-1}$ cm$^{-2}$ arcsec$^{-2}$.
 The \Pa image         
is shallower than the \Hans+[N~II] image, and limits the regions
where the ratio can be measured reliably. Outside the outer marked contours,
the ratio is dominated by noise and background residuals.
The whitest regions correspond to a ratio of 1 or less, while the darkest
areas                  
have a ratio of 10 or more. Most pixels within
the contours of significant \Pa emission have
a greyish shade, corresponding to ratios of about 4. 
}                      
\end{figure}           
\clearpage
\begin{figure}         
\epsscale{1.1}           
\plotone{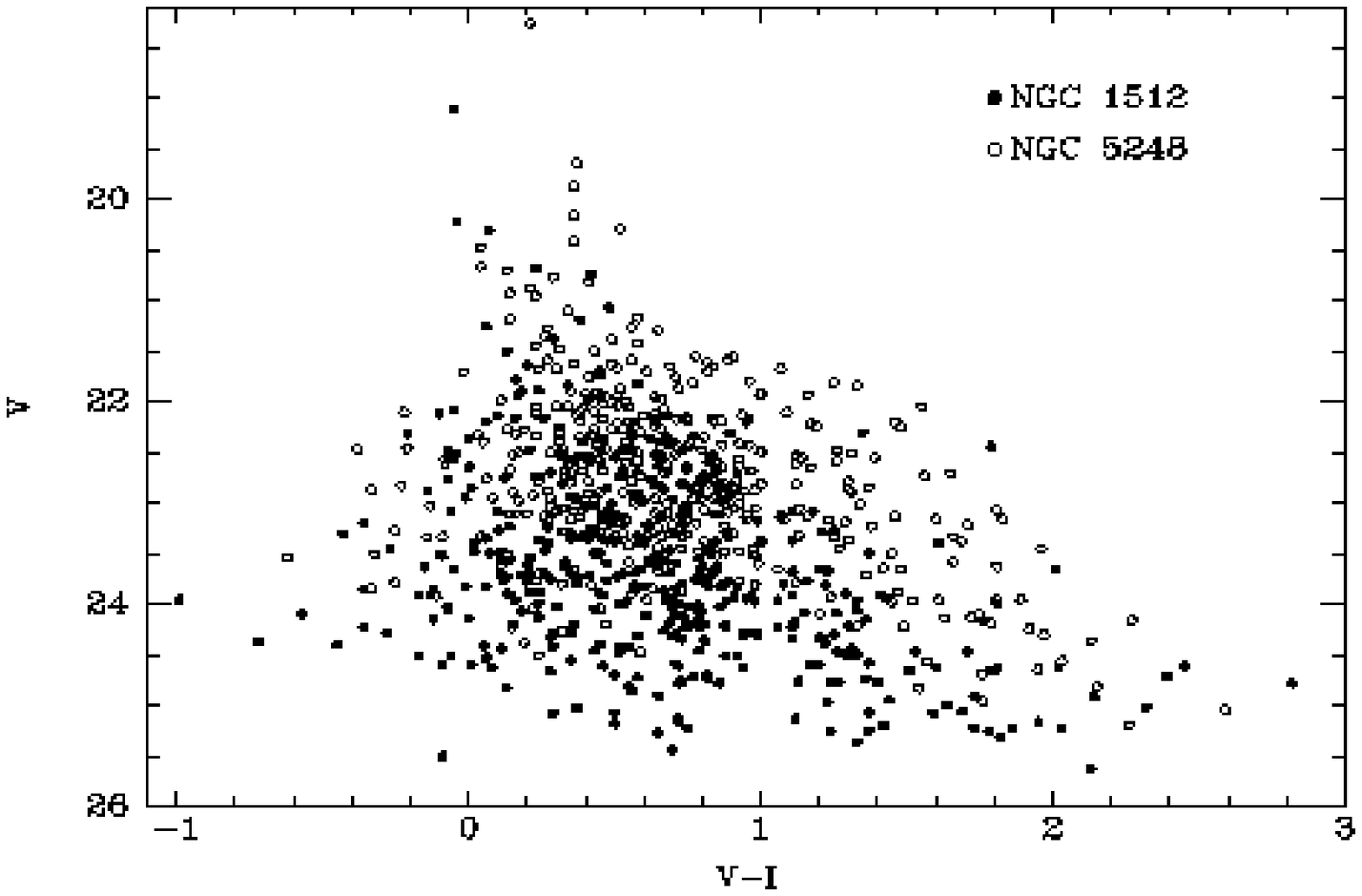}
\caption{
Color-magnitude diagram showing
 all of the sources detected in $V$ and $I$.  
The sources in NGC~5248 have a larger spread in color
than those in NGC~1512, and are typically 0.5-1 mag redder.
}                      
\end{figure}           

\clearpage

\begin{figure}         
\epsscale{1.5}           
\plotone{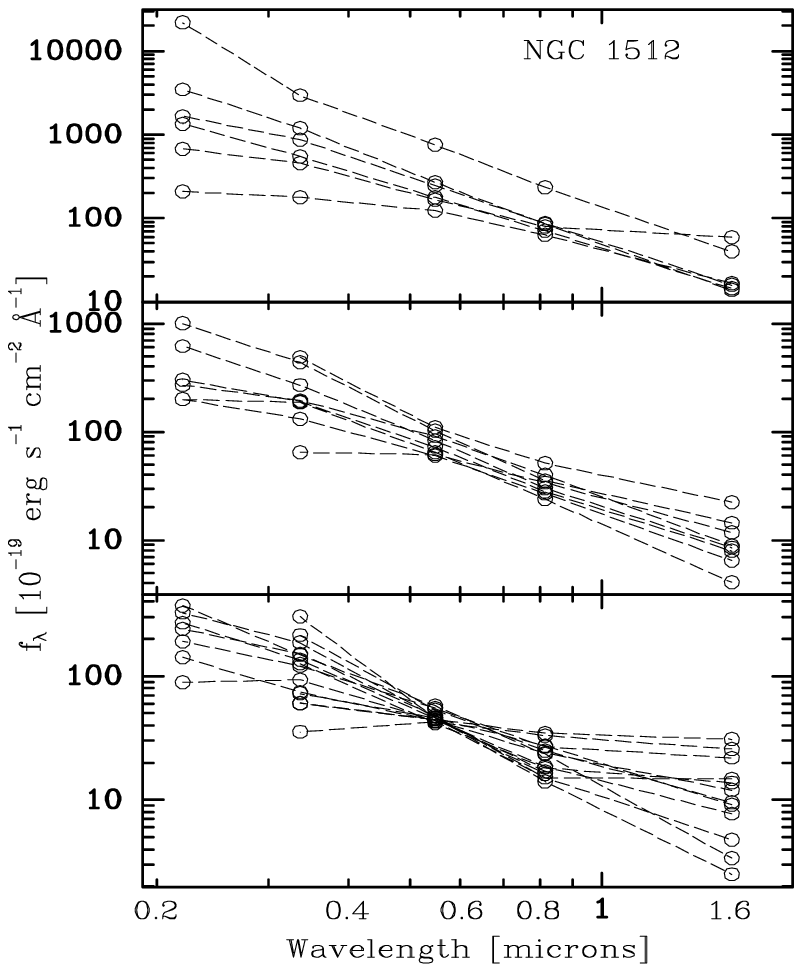}
\caption{
Spectral energy distributions (SEDs), from 0.22 to 1.6  $\mu$m, 
of the 19 clusters that are brightest in $V$ 
in NGC~1512, separated into three panels in decreasing 
brightness groups. Most of the brightest clusters are blue, indicating
an age of a few million years, and little reddening.
Error bars are omitted for clarity. 
}                      
\end{figure}           

\clearpage

\begin{figure}         
\epsscale{1.5}         
\plotone{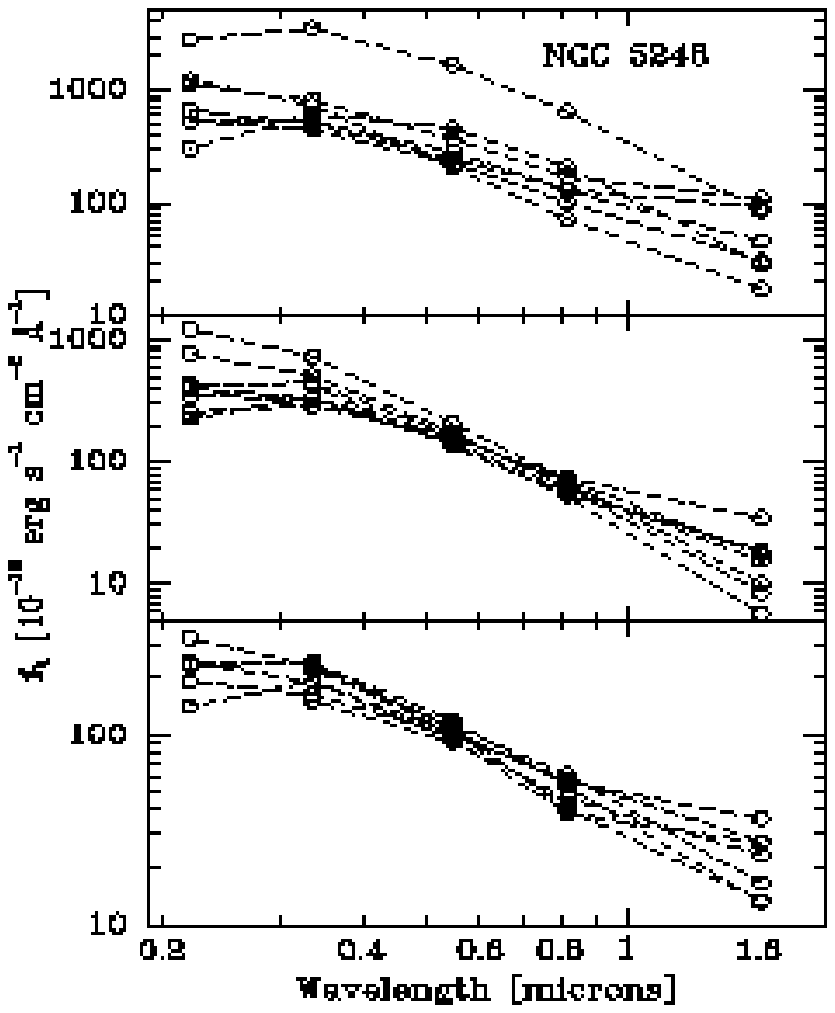}
\caption{Same as Figure 10, but for the 21 clusters brightest in $V$ in 
NGC~5248.}             
\end{figure}

\clearpage

\begin{figure}         
\epsscale{1}           
\plotone{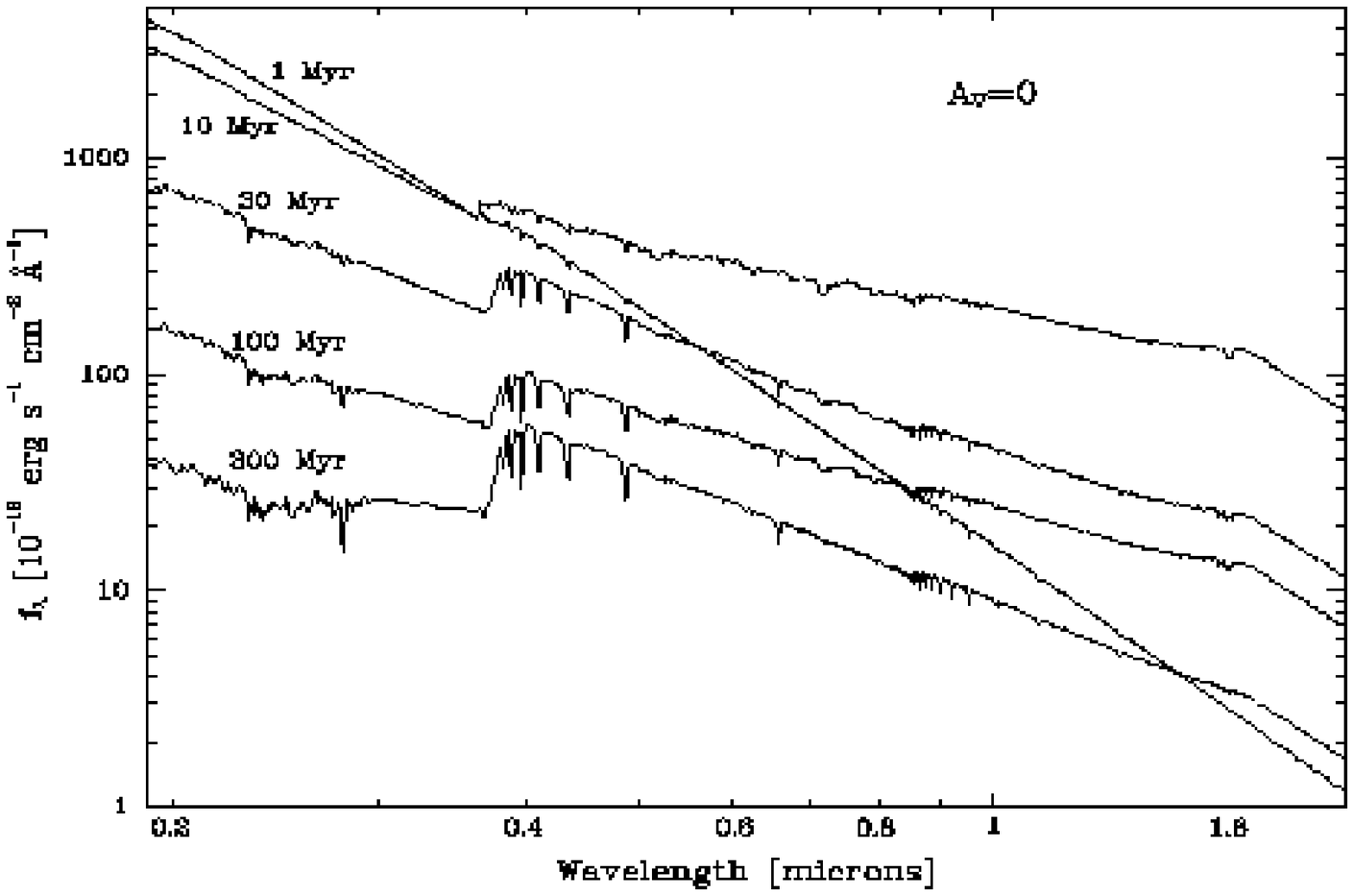}
\caption{
Examples of spectral synthesis models for an unreddened starburst population.
All models assume a Salpeter IMF with upper and lower
mass limits $M_{up}=120  M_{\odot}$ and $M_{low}=1 M_{\odot}$, respectively,
and an exponentially decaying star formation rate with a characteristic
timescale of 
1 Myr.  An arbitrary 
vertical shift has been applied to the models as a whole (but not relative
to each other), corresponding to a scaling with distance-squared or mass.}                      
\end{figure}           
\clearpage
\begin{figure}         
\epsscale{1}           
\plotone{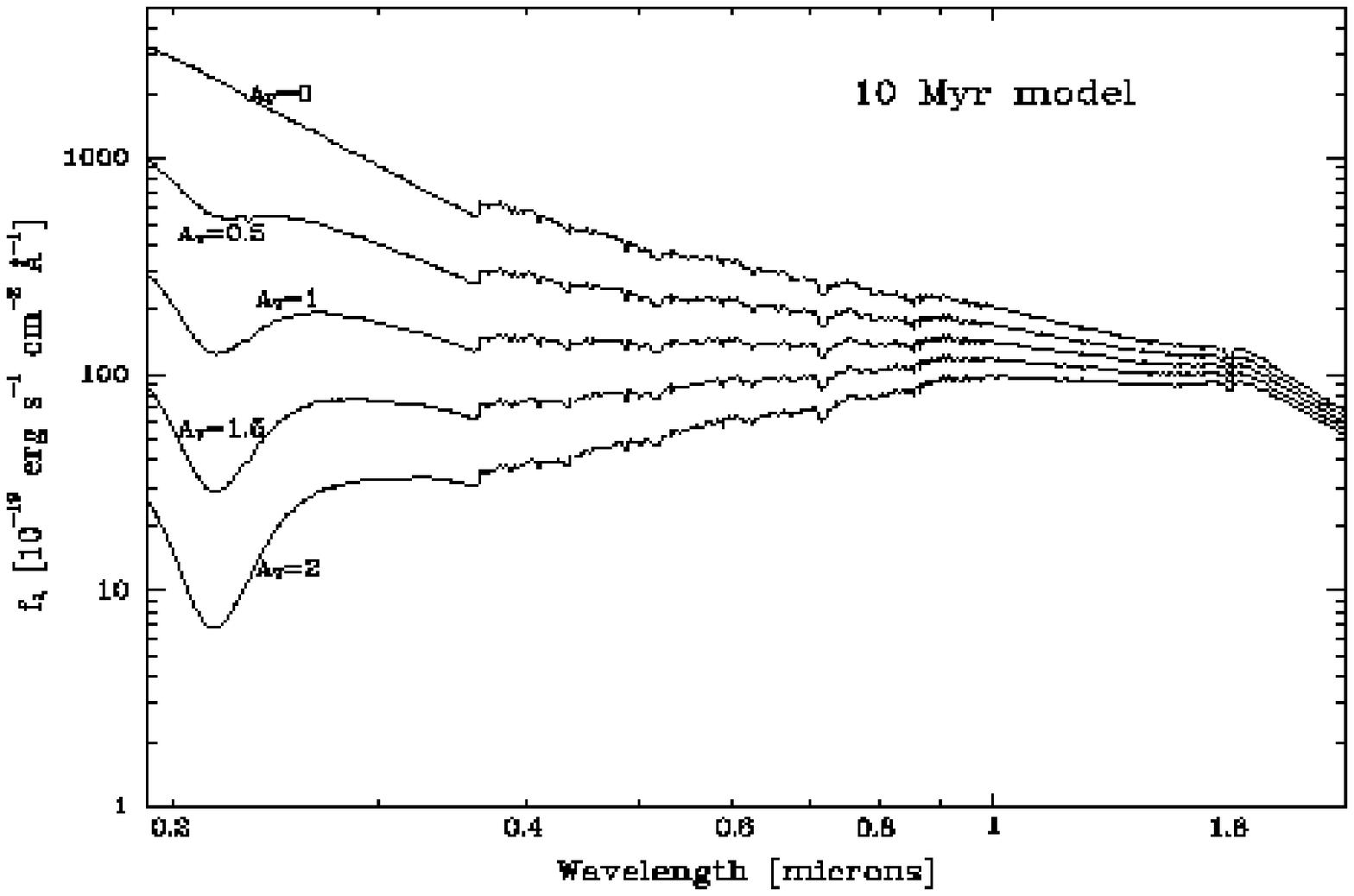}
\caption{Illustration of the effects of foreground extinction on one
of the model SEDs shown in Fig. 12. A Galactic extinction curve
is assumed, with the indicated  magnitudes  of visual extinction.
}                      
\end{figure}           
\clearpage

\begin{figure}         
\plotone{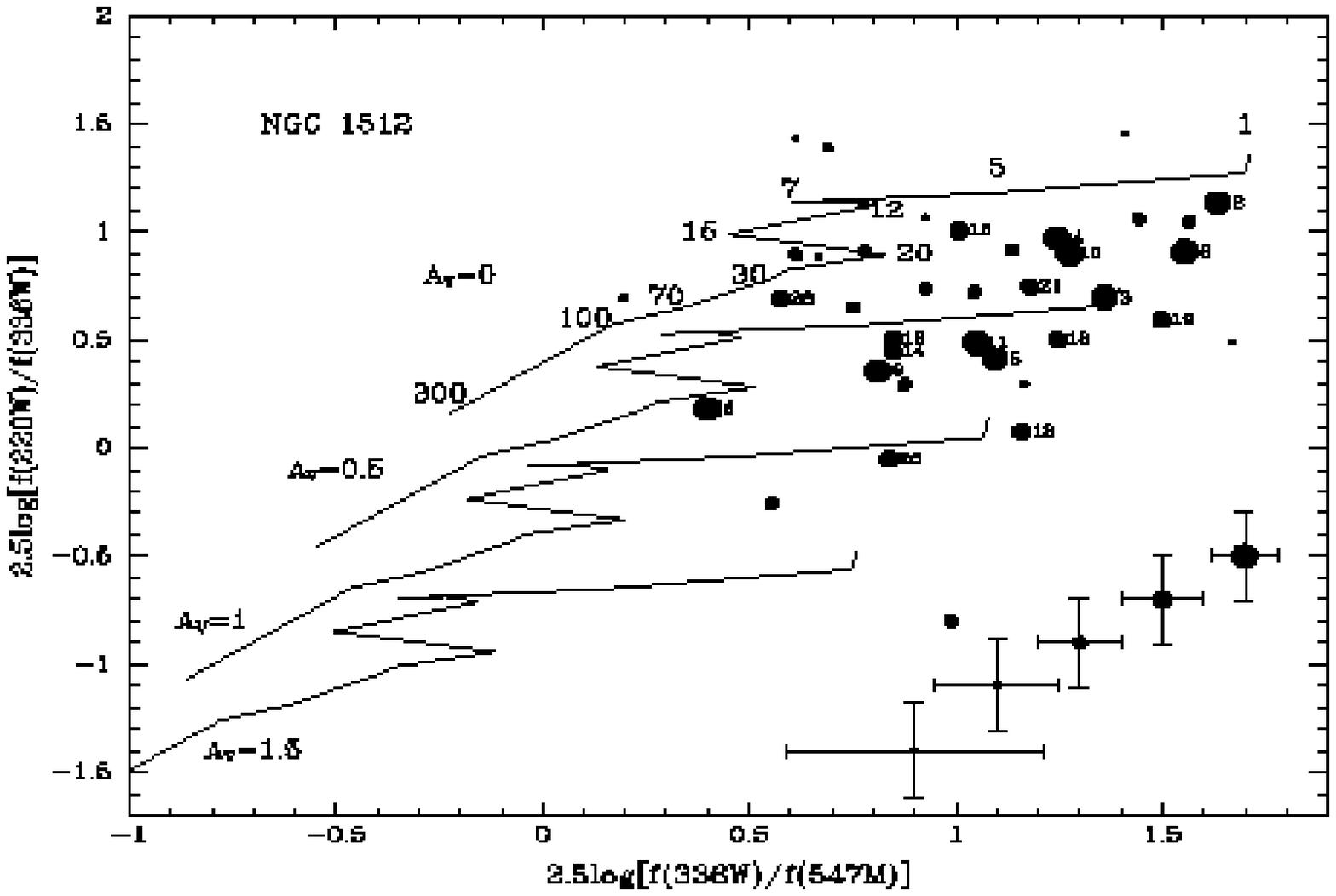}
\caption{Color-color plots of $2.5 \log [f(2300~ {\rm \AA})/f(3350~ {\rm \AA})]$
vs. $2.5 \log [f(3350~ {\rm \AA})/f(5490~ {\rm \AA})]$ (essentially, $U-UV$
vs. $V-U$), for all sources in each galaxy that have a UV measurement.
Point sizes are relative to the $V$ brightness of each source and typical
error bars for the different-sized symbols are displayed in the corner.
The brightest sources are labeled with small numerals according to their
designations in Tables 3 and 4.
  Cluster aging trajectories are shown, based on the spectral synthesis models
 and several values of foreground
 extinction. Ages in Myr are marked along the top trajectory with large
numerals.
The brightest clusters are young,
with ages less than 5~Myr for most of the 15 brightest clusters, and 
less than 10~Myr for most of the UV-detected clusters in general. 
Extinctions are moderate, with the brightest sources clustered
around visual extinctions of $A_V=0.5$ mag in NGC~1512 and $A_V=1$ mag in NGC~5248. 
}             
\end{figure}           
\clearpage

\begin{figure}         
\plotone{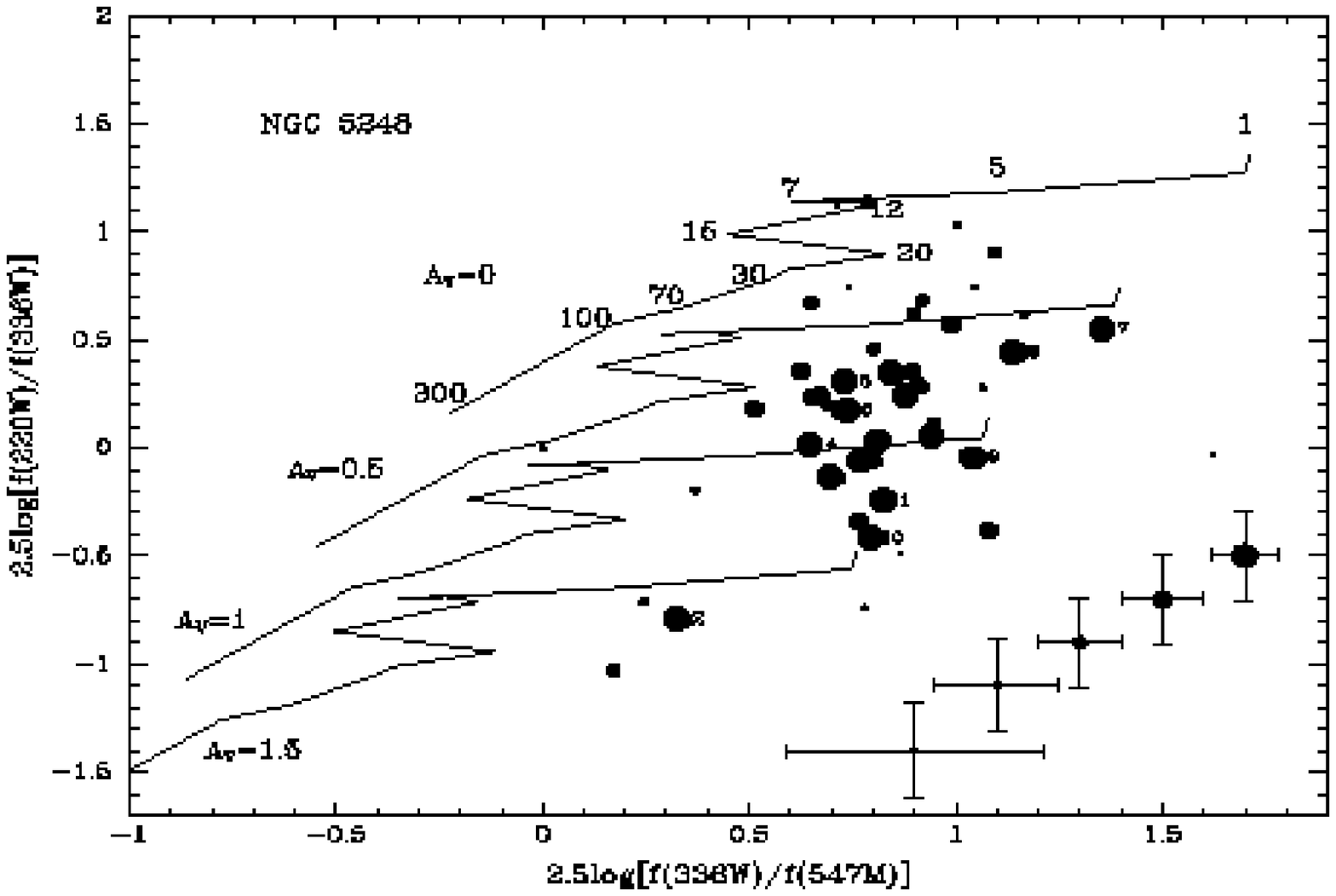}
\end{figure}           
\clearpage

\begin{figure}         
\plotone{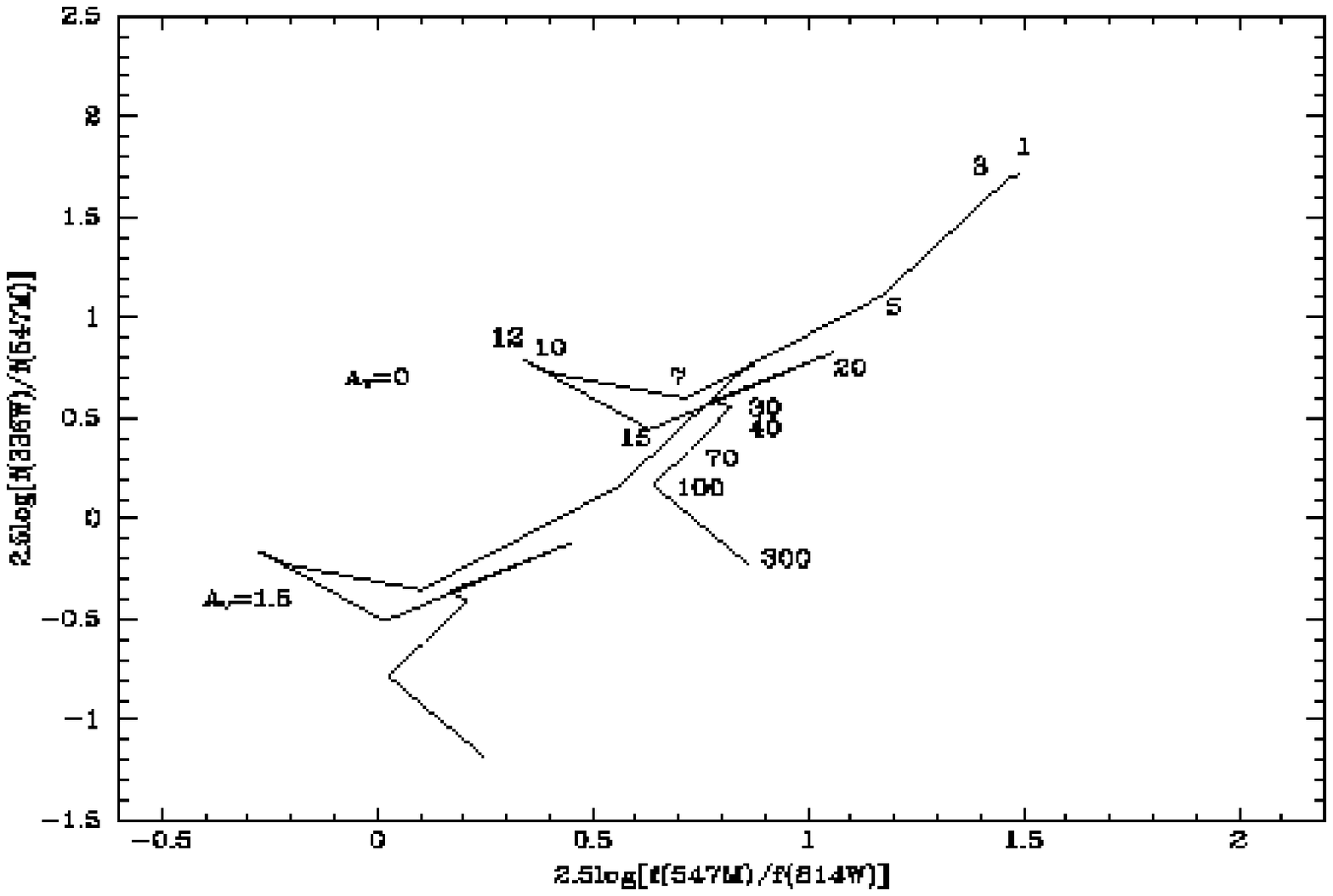}
\caption{
Model age trajectories for two values of extinction in
the plane of
 $2.5 \log [f(3350~ {\rm \AA})/f(5490~ {\rm \AA})]$
vs. $2.5 \log [f(5490~ {\rm \AA})/f(8040~ {\rm \AA})]$ (essentially, $V-U$
and $I-V$, up to additive constants). The ages, in Myr, are marked along the
unreddened trajectory.
The problems of backtracking and age-extinction degeneracy are evident.
} 
\end{figure}            
\clearpage

\begin{figure}         
\plotone{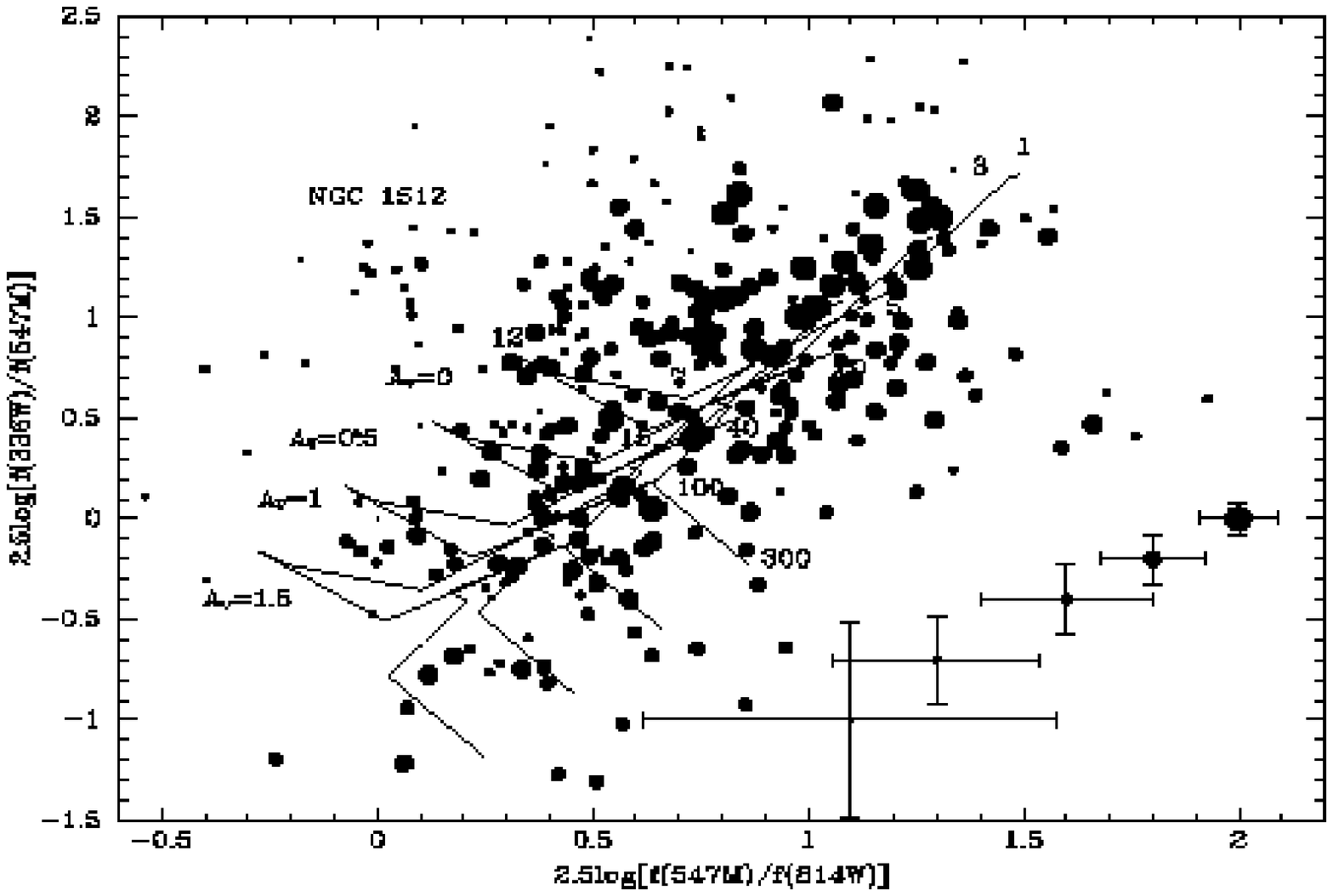}
\caption{
Data for all measured sources in NGC~1512 in
the  $V-U$
 vs. $I-V$ plane of Figure 16. Some age labels are
omitted for clarity. 
The 20 clusters brightest in $V$ 
are shown as the largest points, and each set of the next 100 
brightest sources 
is plotted with progressively
smaller dots. Typical error bars are plotted in the corner.
As in the color-color plot of 
Figure 15, the diagram indicates that many of the bright clusters are
young and only mildly reddened.} 
\end{figure}            
\clearpage
                       
\begin{figure}         
\plotone{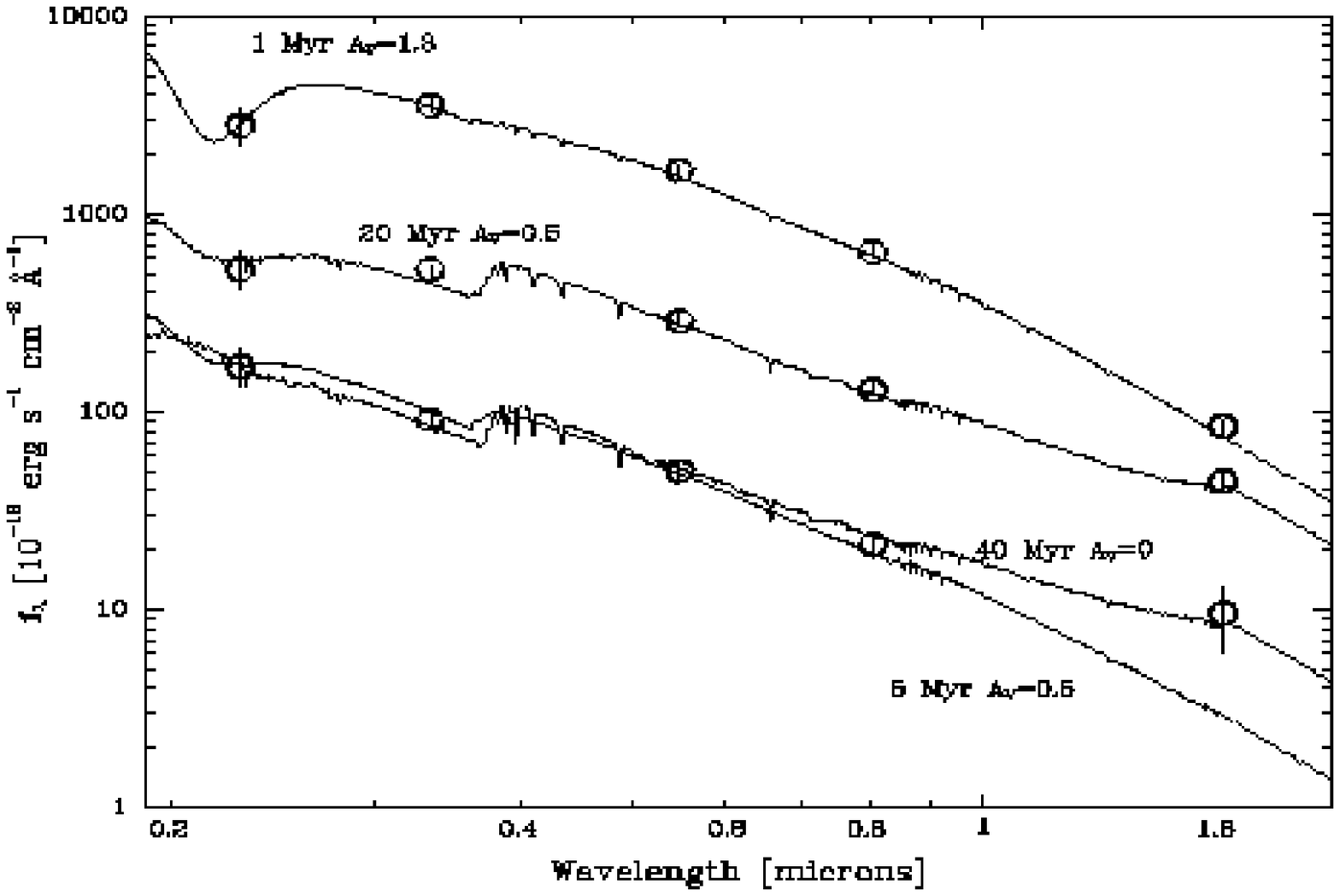}
\caption{Several examples of observed cluster SEDs
in NGC~5248, and their best-fitting models, with ages
and extinctions indicated.
For the fainter cluster, two models are shown --- one that
fits well, and one that is marginally consistent with the data.
} 
\end{figure}           
\clearpage

\begin{figure}          
\plotone{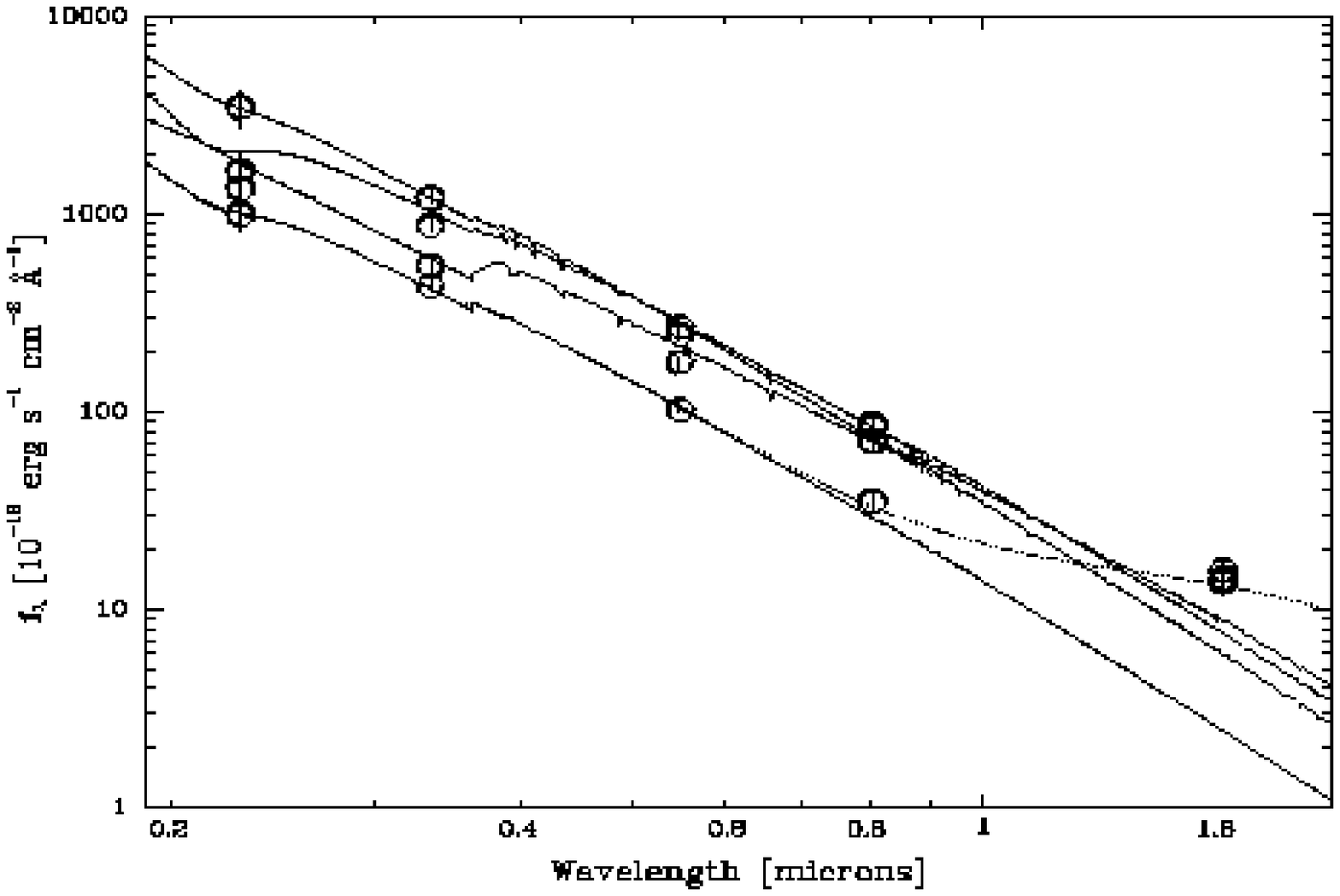}
\caption{Data and
best-fit models for four of the  eight brightest clusters in NGC~1512.
The models range in age from 1 to 5 Myr, and in extinction from
$A_V=0$ to 0.5 mag. In many of the young and bright clusters in 
both galaxies, the data and models are inconsistent at $1.6~ \mu$m.
The dashed line shows the effect of adding to the lower model
a $2000$ K blackbody, scaled to reproduce the excess in the $I$ and IR bands.
If the excess is from dust which is heated by the 
cluster stars to this temperature, the dust must be circumstellar.
} 
\end{figure}            
\clearpage

\begin{figure}          
\plotone{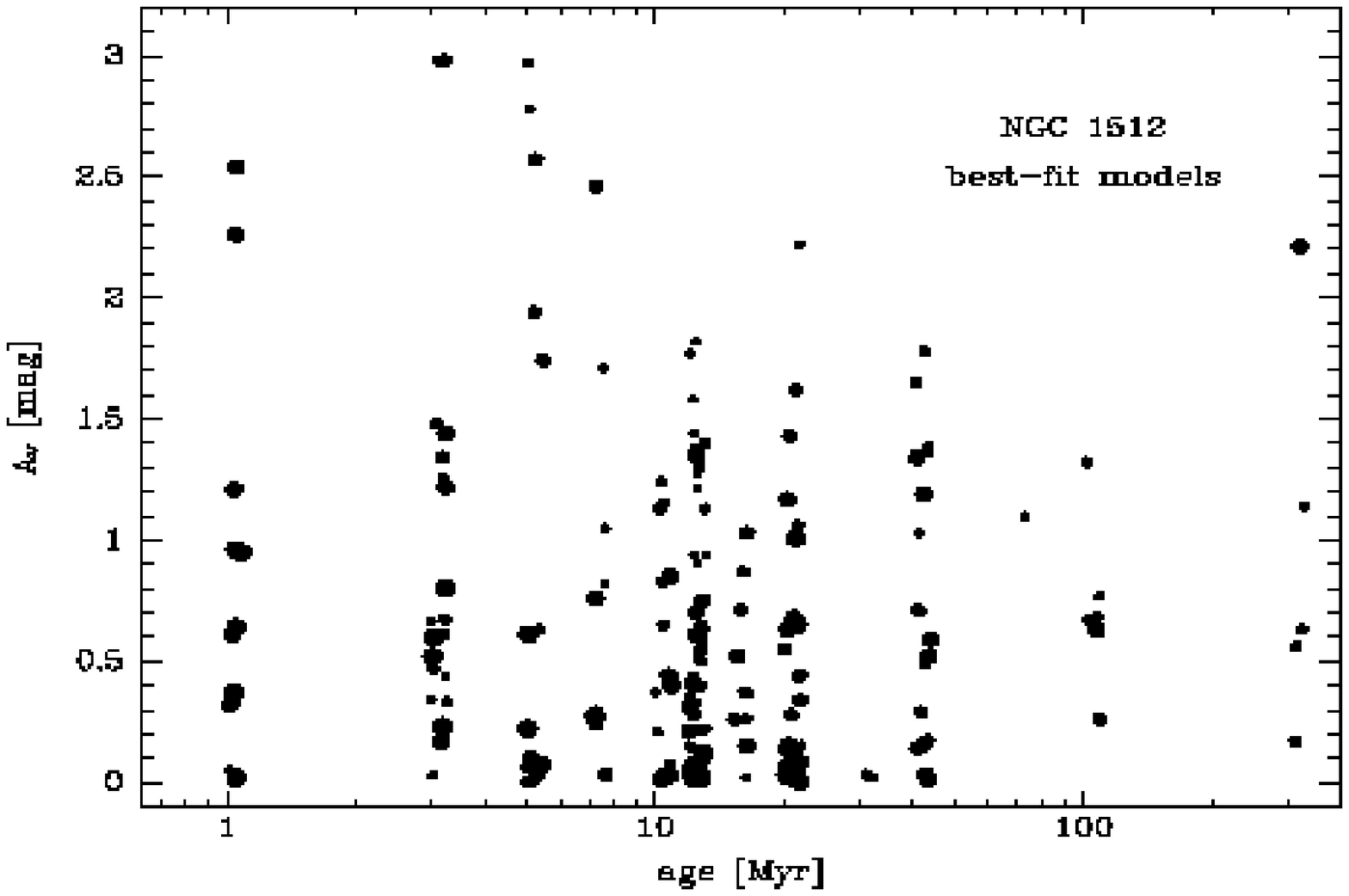}
\caption{Best-fitting ages and extinctions for all sources 
in each galaxy having detections in at least four bands.
As in the previous figures, dot size is proportional
to $V$ brightness. 
For most observed clusters in NGC~1512, 
ages $\ltorder 20$ Myr and extinctions $A_V\ltorder 1$ mag
are indicated, with slightly greater ages and extinctions in NGC~5248.
Some of the plotted best-fit models are
formally inconsistent with the data. In many of these cases, however,
the mismatch is due to an IR excess not predicted by the models, which
still provide a good fit in the other bands.   
} 
\end{figure}            
\clearpage

\begin{figure}          
\plotone{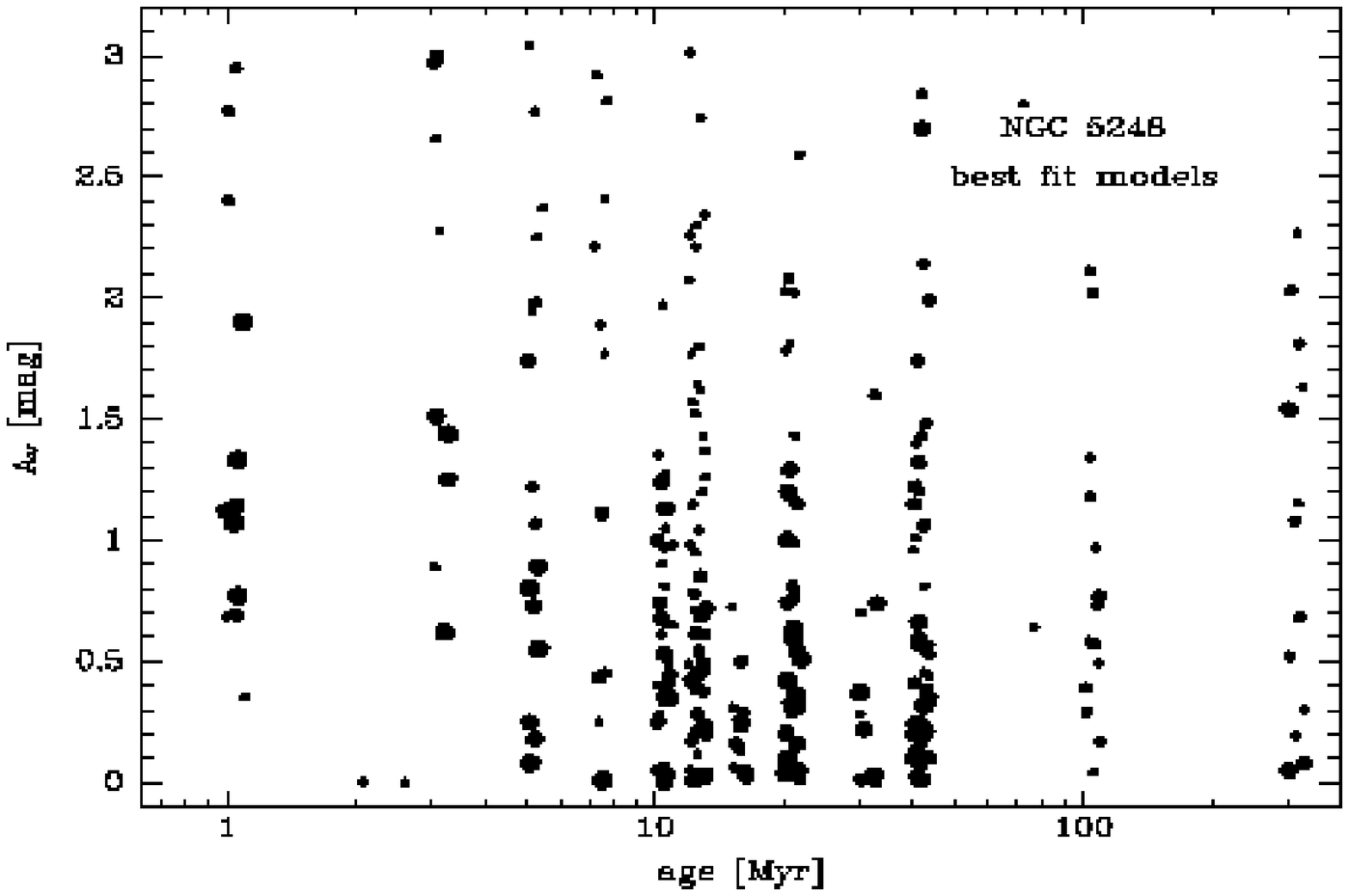}
\end{figure}            

\begin{figure}          
\plotone{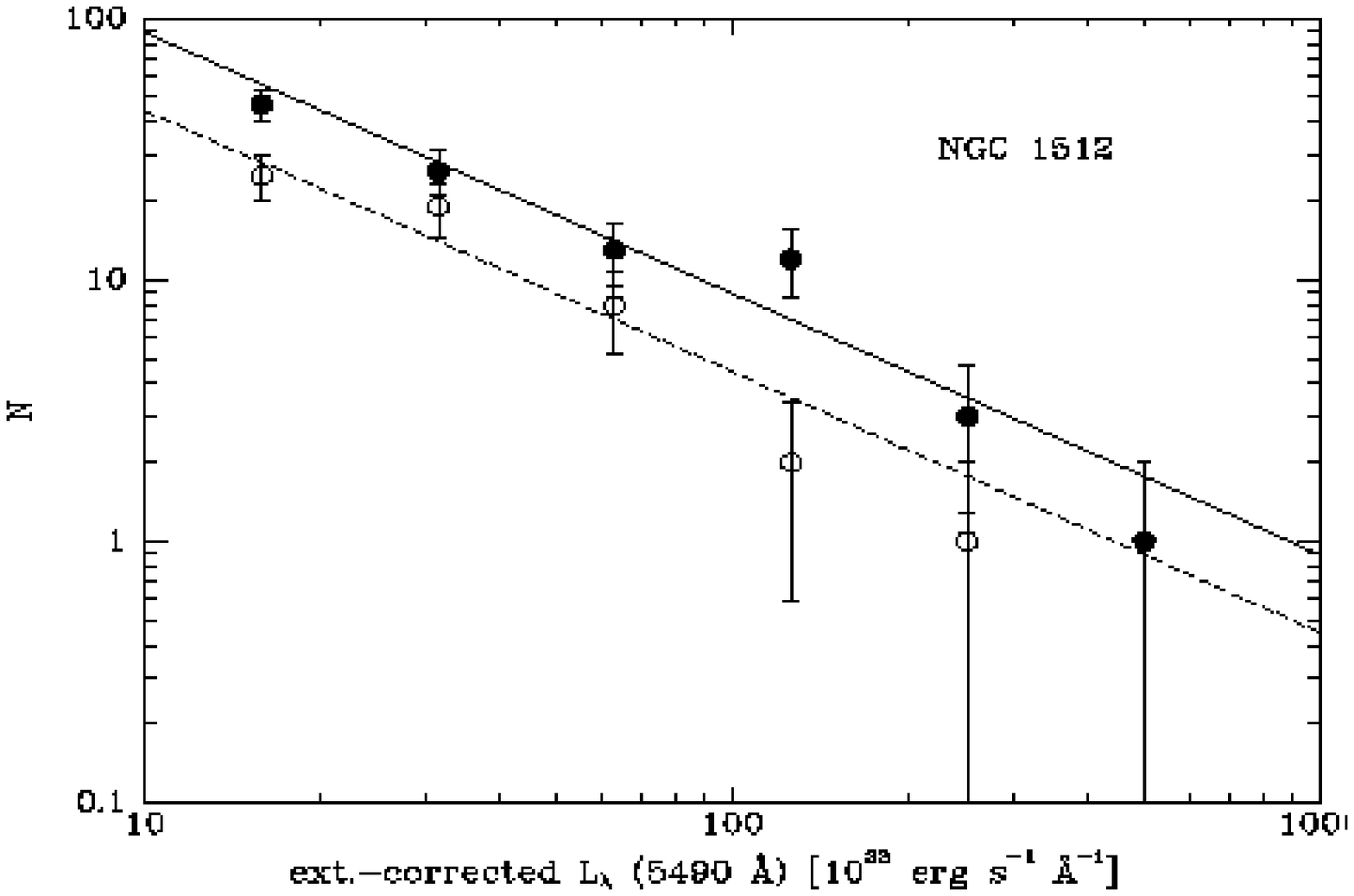}
\caption{Cluster luminosity functions. Filled circles are clusters
whose best-fit models have ages $\le 15$ Myr, and empty circles 
have best-fit models that are older. The $V$ fluxes have been 
corrected for extinction according to each cluster's best-fit model.
Error bars are $1\sigma$ Poisson errors.
The lines are power-law distributions, $N(L) dL \propto L^{-2} dL$,
consistent with the data in the brighter, more complete, bins.
Although at any luminosity, most of the clusters are young, luminosity
evolution shifts the young-cluster distribution horizontally by a factor
of about 6 between ages of 10 Myr and 100 Myr (see Fig. 12). Because of 
the steep distribution, this creates an age bias, where young clusters
are over-represented at every luminosity. After correction for this effect,
the relative fractions of old and young clusters are consistent with 
continuous star formation in the rings, over the last $\sim 100$ Myr.         
} 
\end{figure}            
\clearpage
\begin{figure}          
\plotone{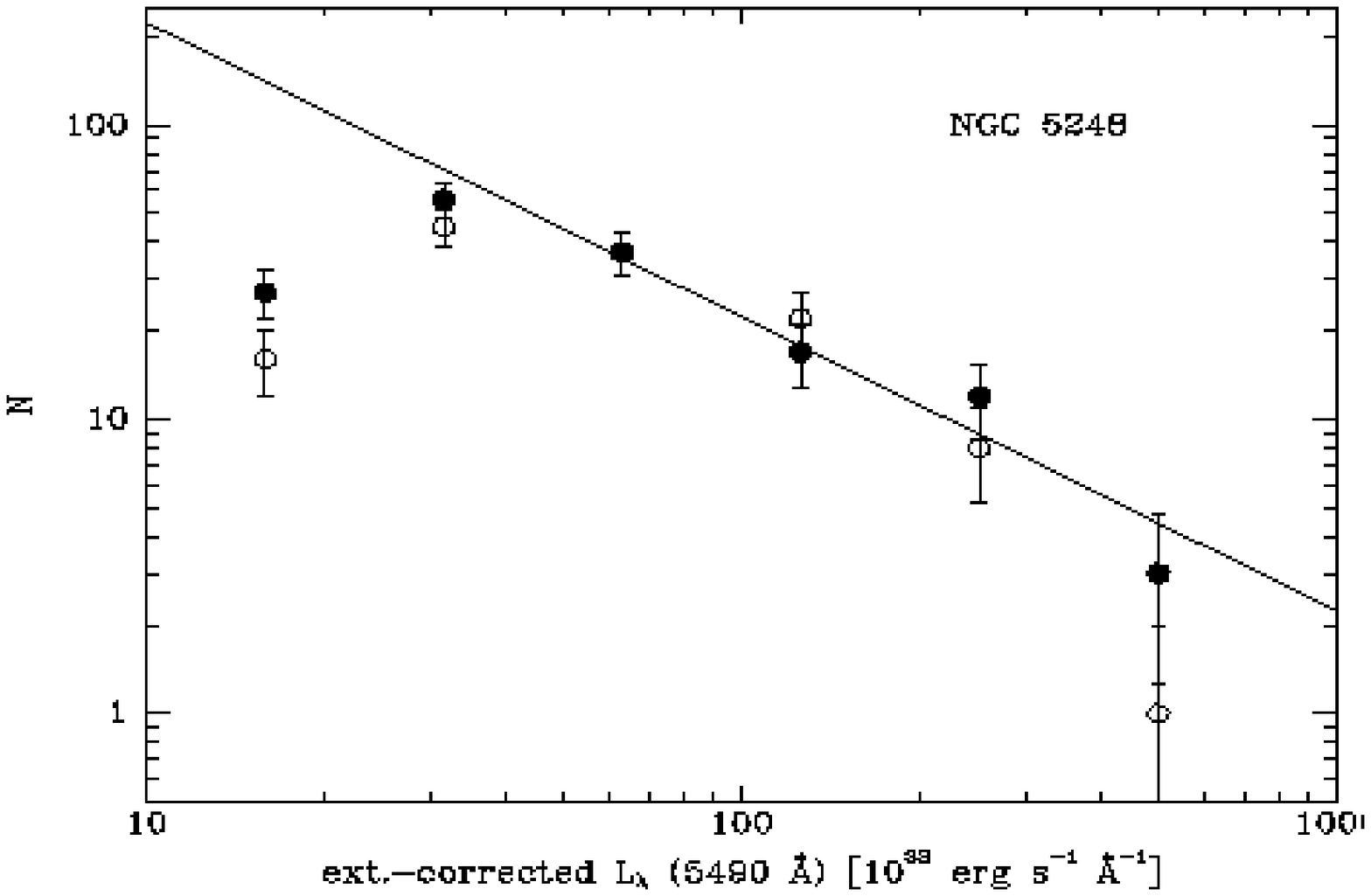}
\end{figure}            

\begin{figure}
\plotone{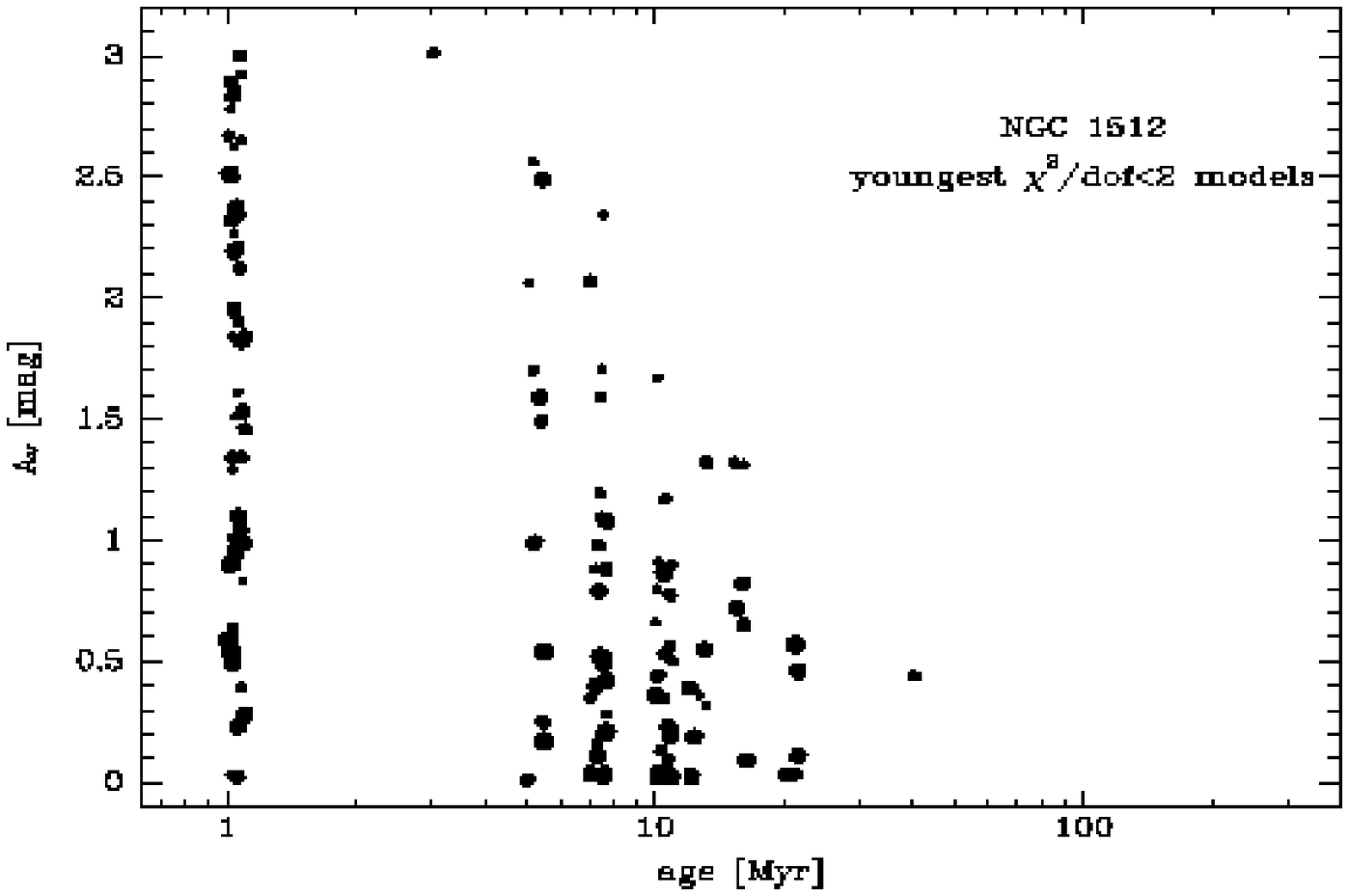}
\caption{Age and extinction of the youngest acceptable 
model that fits each
observed SED at a level of $\chi^2 < 2$ per degree of freedom.
Points are plotted only for the sources   in each galaxy
 that have detections in at least four bands, and satisfy
the above $\chi^2$ criterion, and therefore some clusters do not appear
(e.g., the bright, young clusters with an IR
excess). The figures show that all but a handful of sources can be
consistent with clusters of age $\ltorder 20$ Myr. Thus,
although the best fits (Fig. 20)
suggest that some clusters are old, the presence of such clusters is
not required by the data. The data therefore cannot distinguish between
continuous and episodic star formation in these galaxies.
} 
\end{figure}            

\clearpage
\begin{figure}
\plotone{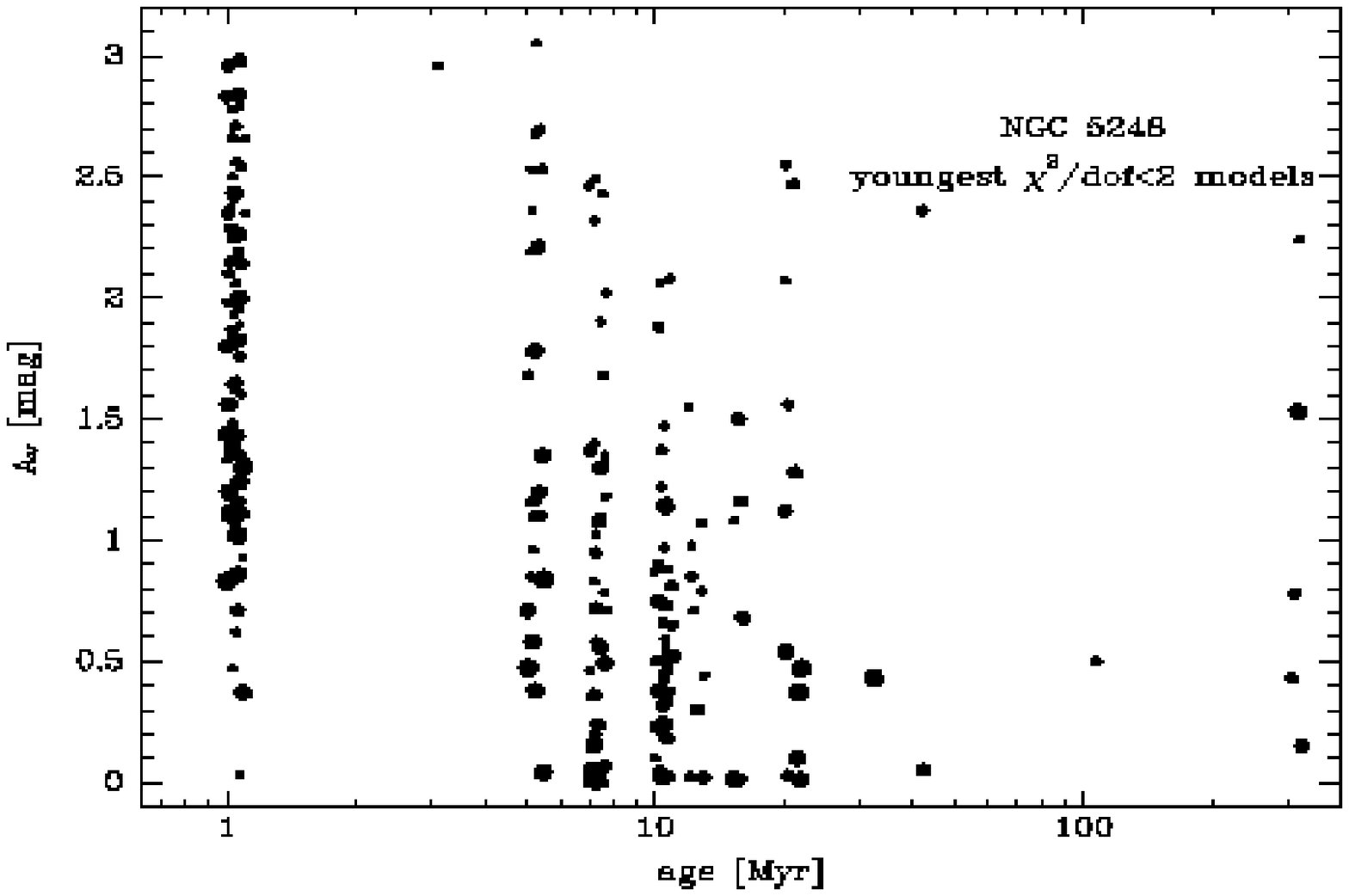}
\end{figure}            

\begin{figure}          
\plotone{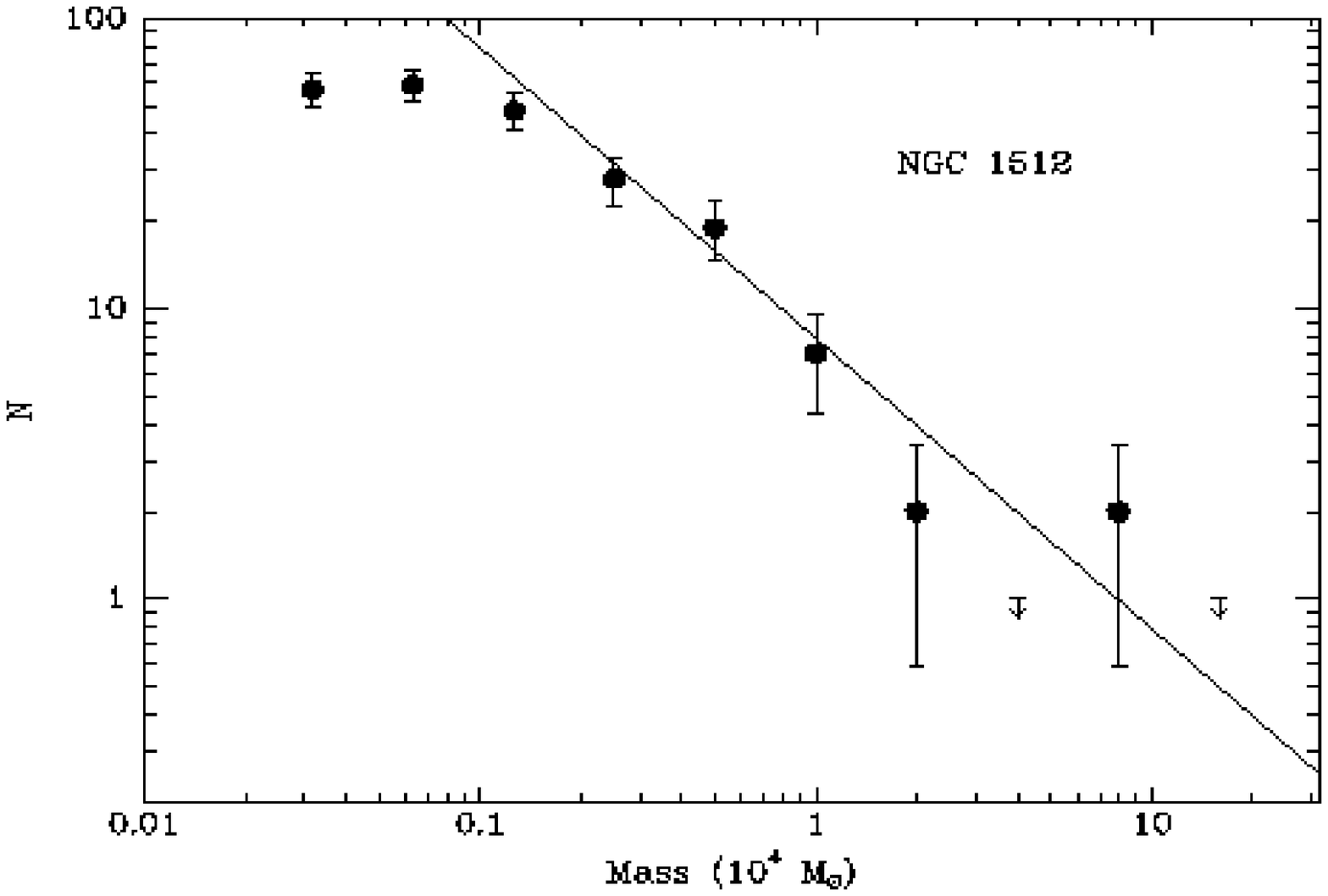}
\caption{Cluster mass functions, based on the best-fit models. 
A $N(m) dm \propto m^{-2} dm$ power law is drawn through the
data for the more massive bins, which suffer less from incompleteness.
The most massive clusters are in the range $10^4$ to $10^5 M_{\odot}$.
The fraction of the most massive clusters is similar to the low fractions
seen in other starburst galaxies having, overall, more clusters.  
} 
\end{figure}            
\clearpage
\begin{figure}          
\plotone{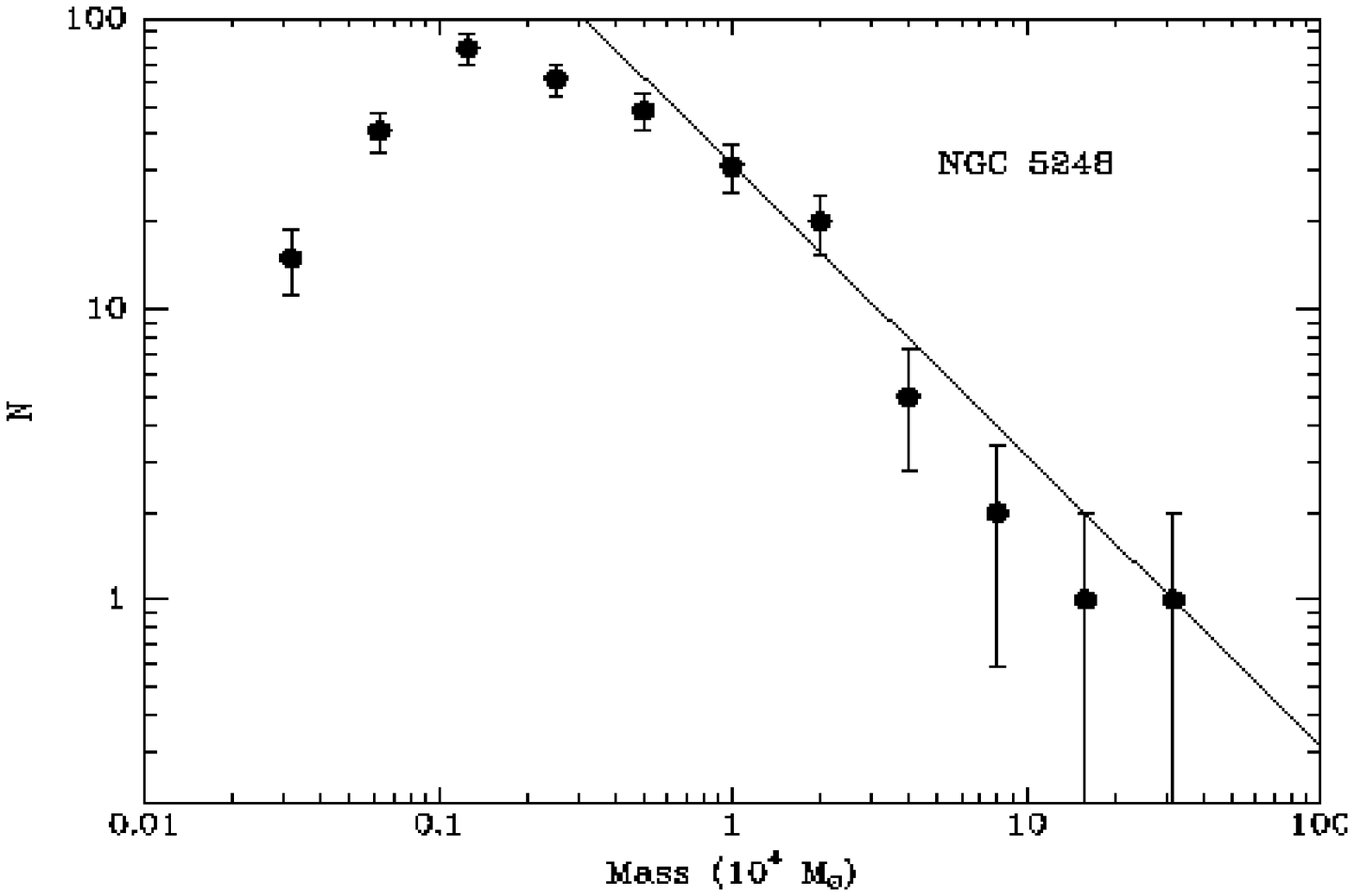}
\end{figure}            
                       
\clearpage

\begin{figure}          
\plotone{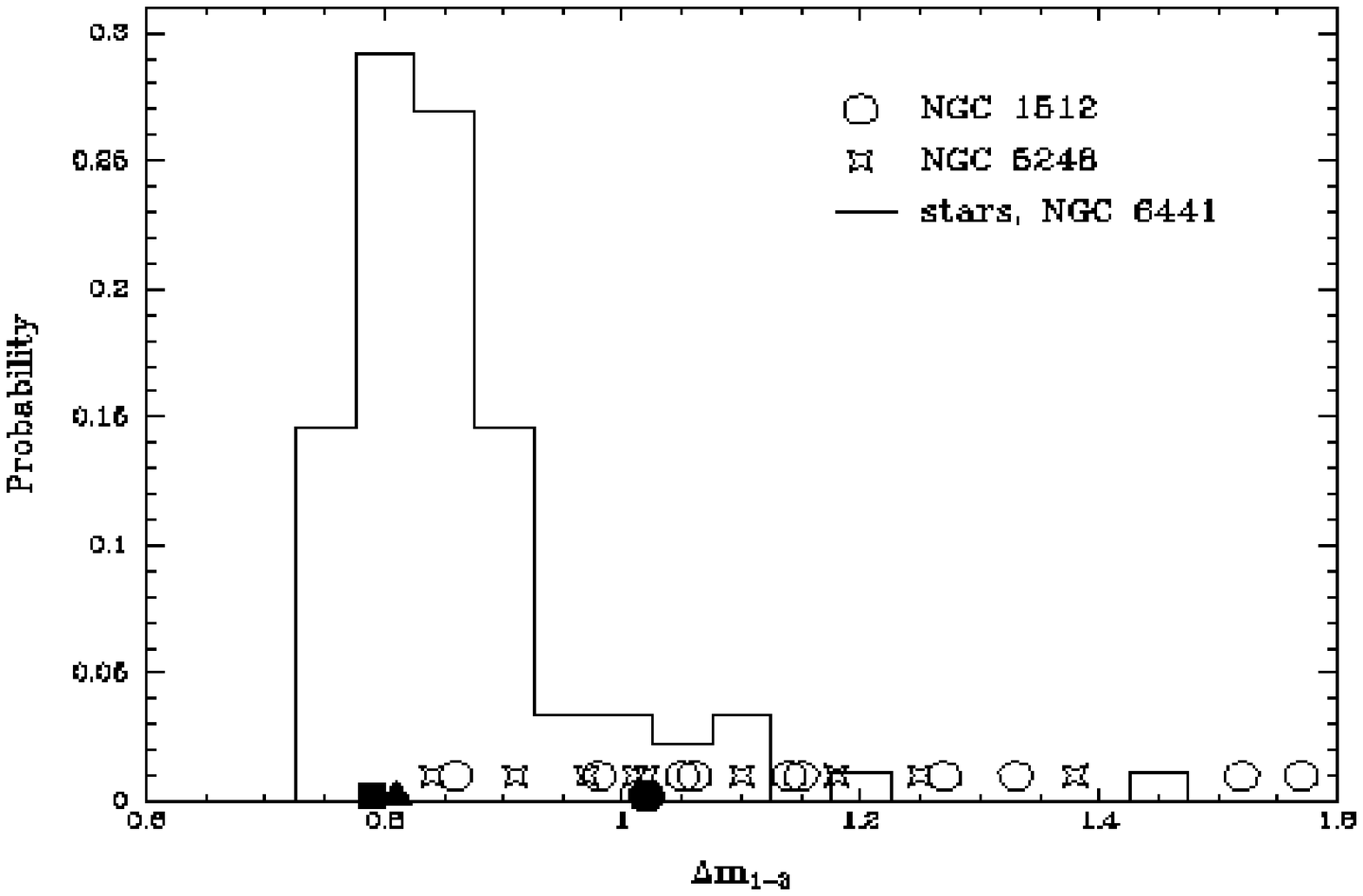}
\caption{
Distribution of $\Delta m_{1-3}$ for stars in a WFPC2 F547M image
of the globular cluster NGC~6441, compared with the values for the 
10 brightest clusters in NGC~1512 (empty circles) and NGC~5248 (star symbols).
Also shown are the two measurements based on Tiny Tim
artificial PSFs, one centered on a pixel center (filled square), and one 
centered on a 4-pixel junction (filled circle). The filled triangle represents
a red, isolated, object in NGC~1512, possibly a foreground star. 
The high-$\Delta m_{1-3}$ tail of the distribution 
for the stars encompasses
many of the SSCs, rendering uncertain the degree to which they are resolved,
but those with $\Delta m_{1-3}>1.1$ are likely resolved.
} 
\end{figure}


\end{document}